\documentclass[showpacs,twocolumn,superscriptaddress]{revtex4}

\usepackage{doi}
\usepackage{hyperref}
\hypersetup{colorlinks=true, linkcolor=blue, citecolor=cyan}
\usepackage{graphicx}
\usepackage{dcolumn}
\usepackage{bm}
\usepackage{xcolor}
\usepackage{amsmath}
\usepackage{amssymb}

\newcommand{\cblue}[1]{{\color{black}#1}}
\newcommand{\cred}[1]{{\color{black}#1}}

\newif\ifizoh
\izohtrue

\begin{document}

\title{Gravitational-Wave Signatures and Periodic Orbits of Charged Particles Around a Five-Dimensional Rotating Black Hole: Comparison with the Kerr Black Hole}

\author{Bahromjon Shokirov}
\email{bahromshokirov05@gmail.com}
\affiliation{National University of Uzbekistan, Tashkent 100174, Uzbekistan}

\author{Tursunali Xamidov}
\email{xamidovtursunali@gmail.com}
\affiliation{Institute for Theoretical Physics and Cosmology, Zhejiang University of Technology, Hangzhou 310023, China}
\affiliation{Institute of Fundamental and Applied Research, National Research University TIIAME, Kori Niyoziy 39, Tashkent 100000, Uzbekistan}
\affiliation{Tashkent State Technical University, 100095 Tashkent, Uzbekistan}

\author{Chen Zhou}
\email{chenzhou@hit.edu.cn}
\affiliation{School of Physics, Harbin Institute of Technology, Harbin 150001, China}

\author{Sanjar Shaymatov}
\email{sanjar@astrin.uz}
\affiliation{Institute of Fundamental and Applied Research, National Research University TIIAME, Kori Niyoziy 39, Tashkent 100000, Uzbekistan}
\affiliation{University of Tashkent for Applied Sciences, Str. Gavhar 1, Tashkent 100149, Uzbekistan}

\author{Chengxun Yuan}
\email{yuancx@hit.edu.cn}
\affiliation{School of Physics, Harbin Institute of Technology, Harbin 150001, China}

\date{\today}

\begin{abstract}

We study periodic orbits of charged particles in the equatorial plane of a five-dimensional ($D=5$) rotating Myers--Perry black hole (BH) immersed in a uniform magnetic field, and the gravitational waves emitted by these orbits in an extreme mass-ratio inspiral. All results are compared with the Kerr BH in the uniform magnetic field. We find that a charged particle is much more weakly bound to the $D=5$ dimensional BH, i.e. its energy at the innermost stable circular orbit is close to unity, $0.95$--$0.99$, compared with $0.78$--$0.94$ for the Kerr BH, and the depth of the potential well is several times smaller. As a result, the periodic orbits of the $D=5$ dimensional BH lie much closer to the separatrix. The magnetic field can make the azimuthal frequency of the orbit smaller than the radial one and produces loops in the orbits far from the BH. Using the numerical kludge method, we compute the gravitational waveforms for a compact object of $10M_\odot$ orbiting a BH of $10^6M_\odot$. At the same magnetic field, the $D=5$ dimensional signal has a longer period, lower frequencies, and an amplitude smaller by about $18\%$. We show that for the magnetic parameter $\beta M=0.02$ the strongest spectral line is at $7.9$\,mHz, compared with $9.9$\,mHz for the Kerr BH. An estimate of the radiation reaction shows that the periodic orbits survive from less than an hour to a few days in the presence of an external magnetic field. Over these lifetimes, the signal-to-noise ratio for LISA is $7$--$31$ for the rotating BH in $D=5$ dimensions, compared to the Kerr BH where it is about $15$--$54$.   

\end{abstract}

\maketitle

\section{Introduction}\label{sec:intro}

Black holes (BHs) are exact solutions of Einstein's equations, and their existence is now supported by gravitational-wave (GW) observations~\cite{Abbott16a,Abbott16b} and by the images of the supermassive BHs in M87* and Sgr~A*~\cite{Akiyama19L1,Akiyama19L6,Akiyama22L12,EHT2022L14}. These observations make BHs a natural place to test gravity in the strong-field regime~\cite{Will14LRR,Psaltis20PRL}. This motivates extending the study of BH physics beyond four-dimensions in general relativity. In particular, higher-dimensional BH solutions are of considerable theoretical interest because extra dimensions naturally arise in several extensions of gravity. Their additional geometric and rotational configurations can produce orbital and gravitational-wave properties that differ from their four-dimensional counterparts. Therefore, investigating particle dynamics and GW signatures around higher-dimensional BHs provides a useful framework for exploring possible imprints of extra dimensions in strong-gravity regimes.

Theories with extra spatial dimensions, motivated by string theory and by brane-world models~\cite{ADD1998,RS1999}, predict BH solutions in more than four dimensions~\cite{EmparanReall2008}. The static BH in $D$ dimensions was found by Tangherlini~\cite{Tangherlini1963}, the rotating BH was found by Myers and Perry~\cite{MyersPerry1986}, and the charged BH solution in five-dimensional Einstein–Maxwell supergravity was developed in Refs.~\cite{Chong05a,Shaymatov20b}. In $D=5$ dimensions, the rotating BH has two independent rotation planes and two spin parameters. An important property of BHs in $D\ge5$ is that test particles have no stable circular orbits~\cite{FrolovStojkovic2003,2022IJMPD..3150120D,Dadhich22a}. The reason is that the gravitational potential falls at least as fast as the centrifugal term, so it cannot hold a particle on a stable orbit, thus affecting the stability of the horizon in higher dimensions \cite{An18,Shaymatov19a,Shaymatov20a,2023Univ....9..190S}. Hence, in higher dimensions (i.e., $D>4$), the rotation-induced repulsive effect of the BH dominates over the attractive contribution of its mass at sufficiently large radial distances \cite{2021PDU....3100758S}. This behavior also occurs for $D=5$ dimensional rotating BHs, which admit no bound orbits of neutral particles at all \cite{2022IJMPD..3150120D,Dadhich22a}.  \cred{In particular, the $D=5$ dimensional Myers–Perry BH generalizes the four-dimensional Kerr solution and introduces additional rotational degrees of freedom, leading to different frame-dragging effects, horizon structures, and orbital properties. $D=5$ dimensional gravity also admits black rings, demonstrating that BH uniqueness and horizon topology differ significantly from their four-dimensional counterparts \cite{EmparanReall2008,Shaymatov20a}. More importantly, the gravitational potential decreases more rapidly with distance in $D=5$ dimensions, resulting in fundamental differences in orbital stability. }

The situation however changes when the BH is placed in an external magnetic field and the particle carries an electric charge. It was shown that a Kerr BH in a uniform magnetic field can be described by a test Maxwell field built from the Killing vectors~\cite{Wald1974}, and this construction was later generalized to the five-dimensional rotating BH (see, e.g.~\cite{AlievFrolov2004,Shaymatov24PRD.110d4042S}. A uniform magnetic field confines the motion of a charged particle across the field lines, and near the BH this changes the stable orbits considerably~\cite{FrolovShoom2010}. Outside the equatorial plane the motion of a charged particle in a weakly magnetized BH spacetime is in general chaotic~\cite{Kovar08,AlZahrani2013}. Along these lines, numerous studies have explored the dynamics of charged particles in weakly magnetized BH spacetimes in a variety of physical scenarios~\cite{FrolovShoom2010,AlievOzdemir2002,Shaymatov15,Tursunov2016,Pavlovic19,Shaymatov19b,Shaymatov21c,Shaymatov23GRG}. Exact solutions describing BHs in a uniform magnetic field have also been found recently~\cite{Podolsky2025KBR}, and their imprints on particle geodesics and GWs have been studied~\cite{2026JCAP...08..035M,LiYanYue2026KBR,Xamidov2026SBR}.

Extreme mass-ratio inspirals (EMRIs) \cite{Hughes_2001,Amaro-Seoane18LRR,Babak17PRD}, in which a stellar-mass compact object orbits a supermassive BH, are important sources of gravitational waves. The properties of these GWs depend strongly on the orbital motion of the smaller body. In particular, bound orbits with zoom--whirl motion are particularly important for understanding particle dynamics in the strong gravitational field near a BH \cite{Glampedakis02PRD,Levin_2008}. Among these trajectories, periodic orbits form a special class in which the particle returns to its initial position after a finite number of radial cycles. For equatorial motion, this occurs when the ratio of the azimuthal frequency $\omega_\varphi$ to the radial frequency $\omega_r$ is rational. Such orbits can be classified by three integers $(z,w,v)$, representing the zoom, whirl, and vertex numbers \cite{Levin_2008,Levin_2009}. This classification has been widely used to investigate orbital dynamics in Schwarzschild, Kerr, and other BH spacetimes \cite{Bambhaniya20,Rana19,JIANG2024,Alloqulov26GW1}. Periodic orbits are also useful for studying the gravitational waveforms and frequency spectra associated with different BH geometries \cite{Yang24JCAP,Shabbir25,Alloqulov25GW,Chen25}. Although radiation reaction causes the orbit to evolve during an actual inspiral, periodic orbits in the conservative approximation provide a useful framework for examining the orbital dynamics and corresponding GW signatures.

It must also be emphasized that EMRIs are among the main sources for space-based detectors LISA~\cite{Amaro-Seoane2017LISA}, Taiji~\cite{10.1093/nsr/nwx116}, and TianQin~\cite{TianQin2016}. Their waveforms carry information about the orbital dynamics and the geometry near the central BH~\cite{Glampedakis02PRD,Gair13CQG,Barack19CQG}. Periodic orbits, classified by three integers $(z,w,v)$, give a convenient way to describe the zoom--whirl structure of such orbits. Their GW signals have been studied for many BH spacetimes~\cite{Barausse14PRD,Cardoso22PRD,Tu23,Yang24JCAP,Shabbir25,Junior24,Haroon25,Lu25GW,Chen25,Li25,Wang25JCAP,Zhang26EPJC,Xamidov2026SBR,Ahmed25GW1,Ahmed25GW2,Alloqulov25GW,Alloqulov26GW1,Sharipov25,JIANG2024,Yang24}. \cred{Considering these studies here make the $D=5$ dimensional rotating BH particularly interesting for investigating the combined effects of spacetime dimensionality, BH rotation, and magnetic fields on orbital dynamics. In this context, comparing the four- and five-dimensional rotating BHs can provide valuable insights into the properties of periodic and zoom–whirl orbits, orbital frequencies, and the resulting GW patterns.}

As stated above, BHs are among the most remarkable predictions of general relativity. In higher-dimensional spacetimes ($D>4$), rotating BH solutions such as the Myers--Perry (MP) BH \cite{MyersPerry1986} possess additional rotational configurations associated with multiple independent rotation parameters. For example, in $D=5$ dimensions, this leads to two independent rotation parameters, which can significantly modify the orbital dynamics around the BH surrounded by an external magnetic field. These distinctive features provide strong motivation for investigating periodic orbits and their associated gravitational-wave signatures in higher-dimensional rotating spacetimes. In this paper, we study periodic orbits of charged particles around the $D=5$ dimensional rotating BH in a uniform magnetic field and the GWs emitted by these orbits. Since in five dimensions the bound orbits can exist only because of the magnetic field \cite{AlievFrolov2004,Shaymatov24PRD.110d4042S}, we compare all results with the same problem for a Kerr BH embedded in a uniform magnetic field. This comparison allows us to separate the effects of the extra dimension from the effects of the magnetic field.

The paper is organized as follows. In Sec.~\ref{sec:spacetime} we describe the $D=5$ dimensional rotating Myers–Perry black hole spacetime and the magnetic field surrounding the BH environment. In Sec.~\ref{sec:motion} we derive the equations of motion of a charged particle and discuss the role of the magnetic field for bound orbits. In Sec.~\ref{sec:periodic}, we introduce the units and parameters to further analyze periodic orbits, which is followed by the main concern of studying the GW model in $D=5$ dimensions and the effects of spin and magnetic field parameters on the GW forms and gravitational radiation reactions in Sec.~\ref{sec:gw}. The conclusions are given in Sec.~\ref{sec:conc}. We use units $G=c=1$, except in Sec.~\ref{sec:gw}.


\section{Five-dimensional rotating black hole and its electromagnetic field}\label{sec:spacetime}

The five-dimensional rotating Myers--Perry BH in Boyer--Lindquist coordinates reads as ~\cite{MyersPerry1986}  
\begin{eqnarray}\label{eq:metric5D}
ds^2&=&-dt^2+\frac{\mu}{\Sigma}\left(dt-a\sin^2\theta\,d\phi-b\cos^2\theta\,d\psi\right)^2\nonumber\\
&+&\frac{\Sigma}{\Delta}\,dr^2+\Sigma\,d\theta^2+(r^2+a^2)\sin^2\theta\,d\phi^2\nonumber\\
&+&(r^2+b^2)\cos^2\theta\,d\psi^2 ,
\end{eqnarray}
where
\begin{eqnarray}\label{eq:SigDel}
\Sigma&=&r^2+a^2\cos^2\theta+b^2\sin^2\theta\, ,\nonumber\\
\Delta&=&\frac{(r^2+a^2)(r^2+b^2)}{r^2}-\mu\, ,
\end{eqnarray}
$\theta\in[0,\pi/2]$, and $\phi,\psi\in[0,2\pi)$. Here $\mu$ is the mass parameter, and $a$ and $b$ are the rotation parameters in the $\phi$ and $\psi$ rotation axes. They are related to the mass $M_5$ and the angular momenta $J_\phi$ and $J_\psi$ of the BH as~\cite{MyersPerry1986}
\begin{equation}
\mu=\frac{8M_5}{3\pi}\, ,\qquad a=\frac{3J_\phi}{2M_5}\, ,\qquad b=\frac{3J_\psi}{2M_5}\, ,
\end{equation}
where the five-dimensional Newton constant is set to one.

In this work, we shall for simplicity consider a five-dimensional MP BH that rotates only in the $\phi$ plane, i.e. $b=0$. Then the horizon is at $r_+=\sqrt{\mu-a^2}$, and the BH exists for $a^2<\mu$ \cite{Shaymatov19a}. For further analysis, we shall consider the motion in the equatorial plane, i.e. $\theta=\pi/2$.

In a vacuum spacetime a Killing vector, taken as a vector potential, solves the source-free Maxwell equations~\cite{Wald1974}. For the five-dimensional rotating BH, the test magnetic field that is uniform at large distances and leaves the BH uncharged is~\cite{AlievFrolov2004}
\begin{equation}\label{eq:AF}
A^{\mu}=\frac{B}{2}\left(\xi^{\mu}_{(\phi)}+a\,\xi^{\mu}_{(t)}\right)+\frac{B_\psi}{2}\left(\xi^{\mu}_{(\psi)}+b\,\xi^{\mu}_{(t)}\right),
\end{equation}
where $\xi_{(t)}=\partial_t$, $\xi_{(\phi)}=\partial_\phi$, and $\xi_{(\psi)}=\partial_\psi$. We set $B_\psi=0$, so that the field lies in the rotation plane of $\phi$. The coefficient $a$ in front of $\xi_{(t)}$
follows from the vanishing electric charge
of the BH, as shown in Ref.~\cite{AlievFrolov2004}.
We choose the constant in $A_t$ in order to have $A_t\to0$ at infinity. Then, for $b=0$, the non-zero components of the vector potential on the equatorial plane are given by 
\begin{equation}\label{eq:A5eq}
A_t=0\, ,\qquad A_\phi=\frac{B}{2}\left(r^2+a^2\right).
\end{equation}
Since $F_{tr}=F_{t\theta}=0$ on this plane, static observers there (outside the ergoregion, $r>\sqrt{\mu}$) measure no electric field. Zero-angular-momentum observers measure the radial electric field $E_r=\omega F_{r\phi}/\alpha$, where $\omega=-g_{t\phi}/g_{\phi\phi}$ and $\alpha$ is the lapse function.

For comparison we use the Kerr BH immersed in a uniform magnetic field described by the Wald solution~\cite{Wald1974},
\begin{equation}\label{eq:Wald}
A^{\mu}=\frac{B}{2}\left(\xi^{\mu}_{(\phi)}+2a\,\xi^{\mu}_{(t)}\right).
\end{equation}
The motion of charged particles in such a field has been studied in detail~\cite{AlievOzdemir2002,AlZahrani2013,Shaymatov15,Tursunov2016}. 
As in five dimensions, we choose the field
for which the BH has no electric charge
and the gauge $A_t\to0$ at infinity. On the equatorial plane both BH spacetimes satisfy $g_{rr}=r^2/\Delta$ and $g_{t\phi}^2-g_{tt}g_{\phi\phi}=\Delta$, with $\Delta=r^2-2Mr+a^2$ for Kerr, so that the equations of motion below hold for both. We compare the two BHs at the same values of the dimensionless spin and of the magnetic field parameter, as we further discuss details in Sec.~\ref{sec:periodic}.

\section{Motion of charged particles}\label{sec:motion}

The motion of a particle with mass $m$ and charge $q$ is described by the Lagrangian \cite{Misner73}
\begin{equation}
\mathcal{L}=\frac{1}{2}\,g_{\mu\nu}\dot x^\mu\dot x^\nu+\frac{q}{m}A_\mu\dot x^\mu\, ,
\end{equation}
where the dot denotes the derivative with respect to the proper time $\tau$. Since the metric and $A_\mu$ do not depend on $t$ and $\phi$, the energy and the angular momentum per unit mass are conserved ($E\to E/m$ and $L\to L/m$),
\begin{eqnarray}\label{eq:cons}
g_{tt}\dot t+g_{t\phi}\dot\phi+\frac{q}{m}A_t&=&-E\, ,\nonumber\\
g_{t\phi}\dot t+g_{\phi\phi}\dot\phi+\frac{q}{m}A_\phi&=&L\, .
\end{eqnarray}
We introduce the magnetic parameter $\beta$ and, for convenience, the notation $\tilde E$ and $\tilde L$,
\begin{equation}\label{eq:beta}
\beta=\frac{qB}{2m}\, ,\qquad \tilde E=E+\frac{q}{m}A_t\, ,\qquad \tilde L=L-\frac{q}{m}A_\phi\, .
\end{equation}
Using the vector potential~(\ref{eq:A5eq}), these quantities can be written in the equatorial plane of the five-dimensional BH in the following form:
\begin{equation}\label{eq:EL5}
\tilde E=E\, ,\qquad \tilde L=L-\beta\,(r^2+a^2)\, .
\end{equation}
In the same way, for the Kerr BH we have $\tilde E=E+2a\beta M/r$ and $\tilde L=L-\beta\,(r^2+a^2-2Ma^2/r)$. Solving Eq.~(\ref{eq:cons}) for $\dot t$ and $\dot\phi$ and using the relation $g_{t\phi}^2-g_{tt}g_{\phi\phi}=\Delta$, we obtain following expressions for the five-dimensional BH
\begin{eqnarray}\label{eq:dot5}
\dot t&=&\frac{(r^4+a^2r^2+\mu a^2)\tilde E-a\mu\tilde L}{r^2\Delta}\, ,\nonumber\\
\dot\phi&=&\frac{a\mu\tilde E+(r^2-\mu)\tilde L}{r^2\Delta}\, .
\end{eqnarray}

Substituting $\dot t$ and $\dot\phi$ into the normalization condition $g_{\mu\nu}\dot x^\mu\dot x^\nu=-1$ and using $g_{rr}=r^2/\Delta$, we obtain the radial equation
\begin{equation}\label{eq:rdot}
r^4\dot r^2=\mathcal{R}(r)\equiv r^2\left(g_{\phi\phi}\tilde E^2+2g_{t\phi}\tilde E\tilde L+g_{tt}\tilde L^2-\Delta\right).
\end{equation}
For the BHs in $D=4,5$ dimensions, the function $\mathcal{R}$ takes the form
\begin{eqnarray}
\text{5D:}\quad \mathcal{R}&=&(\tilde E^2-1)\,r^4+\left(\mu-a^2+a^2\tilde E^2-\tilde L^2\right)r^2\nonumber\\
&+&\mu\,(\tilde L-a\tilde E)^2\, ,\label{eq:R5}\\
\text{Kerr:}\quad \mathcal{R}&=&(\tilde E^2-1)\,r^4+2Mr^3+\left(a^2\tilde E^2-a^2-\tilde L^2\right)r^2\nonumber\\
&+&2Mr\,(\tilde L-a\tilde E)^2\, .\label{eq:R4}
\end{eqnarray}
The comparison of these expressions shows the main difference between the two BHs: the Newtonian term $2Mr^3$ of the Kerr BH is replaced in five dimensions by $\mu r^2$, because the gravitational potential falls as $1/r^2$ instead of $1/r$ \cite{2022IJMPD..3150120D,Dadhich22a}.

The function $\mathcal{R}$ in Eq.~(\ref{eq:rdot}) is quadratic in $\tilde E$. Solving the condition $\mathcal{R}=0$ for $\tilde E$, taking the root with $\dot t>0$, and using $\tilde E=E+(q/m)A_t$, we find that the turning points of the radial motion are given by $E=V_{\rm eff}(r)$, where \cite{2022IJMPD..3150120D,Shaymatov24PRD.110d4042S}
\begin{equation}\label{eq:Veff}
V_{\rm eff}=-\frac{q}{m}A_t+\omega\tilde L+\sqrt{\frac{\Delta}{g_{\phi\phi}}\left(1+\frac{\tilde L^2}{g_{\phi\phi}}\right)}
\end{equation}
and $\omega=-g_{t\phi}/g_{\phi\phi}$. Since $\tilde L$ does not depend on $E$, the effective potential depends only on $r$ and $L$, and the motion is allowed where $E\ge V_{\rm eff}$. For the five-dimensional BH, $(q/m)A_t=0$ and $\omega=a\mu/(r^4+a^2r^2+\mu a^2)$. For the numerical integration, we use Eq.~(\ref{eq:dot5}) together with the second-order form of the radial equation,
\begin{equation}\label{eq:rddot}
\frac{\mathcal{R}(r)}{r^4}=\dot r^2\, ,\qquad \Psi(r)\equiv\ddot r=\frac{1}{2}\,\frac{d\Psi}{dr}\, ,
\end{equation}
which has no difficulty at the turning points.
\begin{figure*}[t]
\centering
\includegraphics[width=0.85\textwidth]{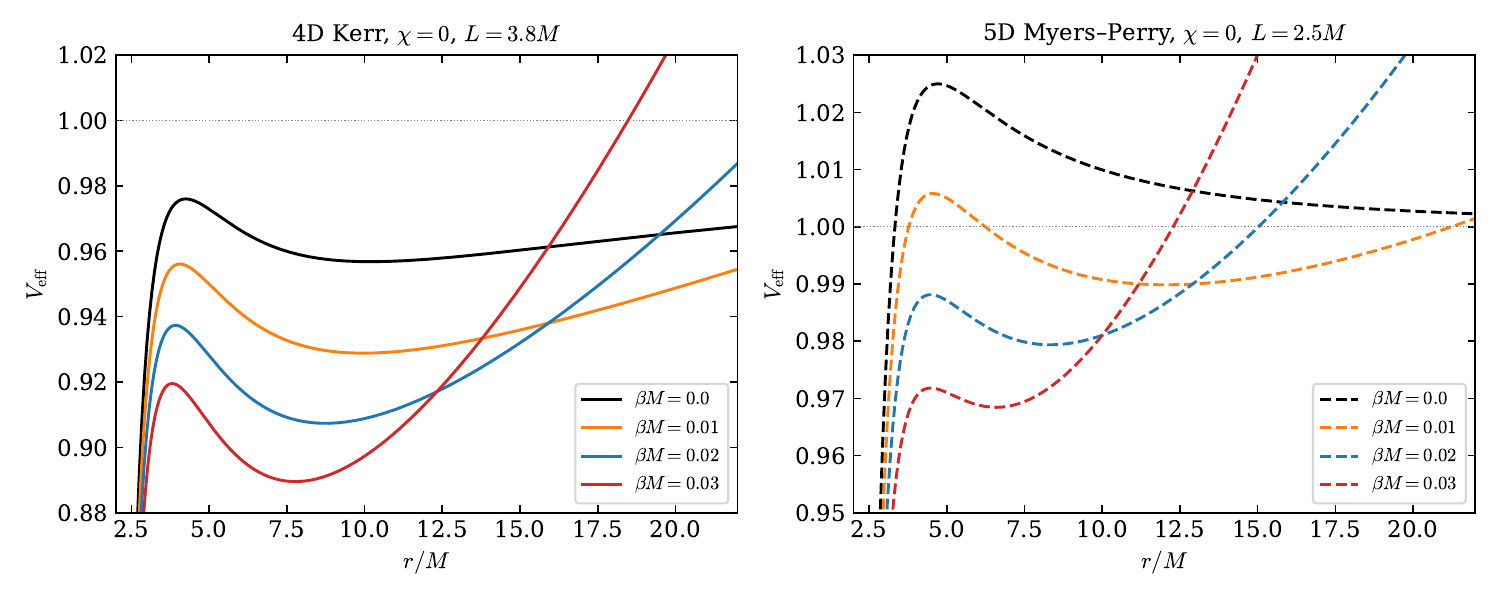}
\caption{Effective potential $V_{\rm eff}(r)$ for $\chi=0$. Left: Kerr BH with $L=3.8M$; right: five-dimensional Myers--Perry BH with $L=2.5M$. 
The gray dotted line shows $V_{\rm eff}=1$.}
\label{fig:veff}
\end{figure*}

The radial component of the Lorentz force per unit mass is $f_r=(q/m)F_{r\nu}\dot x^\nu$, where $F_{\mu\nu}=\partial_\mu A_\nu-\partial_\nu A_\mu$. For the five-dimensional BH, Eq.~(\ref{eq:A5eq}) gives $F_{rt}=0$ and $F_{r\phi}=Br$ on the equatorial plane, so that $f_r=2\beta r\dot\phi$. Hence the Lorentz force points away from the BH for $\beta\dot\phi>0$ and toward it for $\beta\dot\phi<0$. For the Kerr BH the same holds for the magnetic part of the force, since $F_{r\phi}=B(r+Ma^2/r^2)>0$. Here, we consider $\beta>0$ and $L>0$, i.e., the case in which the Lorentz force points away from the BH when the particle moves in the direction of the BH rotation (anti-Larmor motion~\cite{AlievOzdemir2002,Tursunov2016}).

In Fig.~\ref{fig:veff}, we show the effective potential for $\chi=0$ and various values of $\beta$. For the five-dimensional BH without magnetic field (black dashed line) the potential has only a maximum and decreases monotonically at larger $r$. Therefore, there is no stable circular orbit.
When the magnetic field is switched on, the potential grows at large $r$ for both BHs. For the Kerr BH, which has a potential minimum/well already for $\beta=0$, the magnetic field makes the well deeper and moves it to smaller radii and energies. For the five-dimensional BH, the growth of the potential at large $r$ creates the potential well, but its depth is much smaller than in the Kerr case.

\subsection{Bound orbits and the role of the magnetic field in $D=5$ dimensions}\label{sec:bound}

Let us first consider the five-dimensional BH without magnetic field, $\beta=0$. In this case $\tilde E=E$ and $\tilde L=L$, and introducing the variable $x=r^2$, we can write Eq.~(\ref{eq:R5}) as a quadratic function of $x$,
\begin{eqnarray}
\mathcal{R}&=&(E^2-1)x^2+\left(\mu-a^2+a^2E^2-L^2\right)x\nonumber\\
&+&\mu\,(L-aE)^2\, .
\end{eqnarray}
For $E<1$ the coefficient of $x^2$ is negative, while the free term $\mu(L-aE)^2$ is non-negative. Therefore, the product of the roots is non-positive, and $\mathcal{R}$ has at most one positive root. This means that for $E<1$ there is no inner turning point outside the horizon, and the particle falls into the BH. For $E\ge1$ the particle either escapes to infinity or falls into the BH. Thus, without the magnetic field there are no bound orbits in the equatorial plane of the five-dimensional BH, in agreement with the absence of stable circular orbits found in Refs.~\cite{FrolovStojkovic2003,2022IJMPD..3150120D,Dadhich22a}.

The reason becomes clear if we compare Eq.~(\ref{eq:R5}) with Eq.~(\ref{eq:R4}). For the Kerr BH, the function $\mathcal{R}$ contains the term $2Mr^3$, which comes from the Newtonian attraction and makes bound orbits possible already for $\beta=0$. In $D=5$ dimensions, this term is replaced by $\mu r^2$, because the gravitational potential falls as $1/r^2$ instead of $1/r$ \cite{2021PDU....3100758S}. In the Newtonian limit with $a=0$, the effective potential per unit mass is then $U=(L^2-\mu)/(2r^2)$, i.e., the gravitational and centrifugal terms have the same dependence on $r$, and their sum has no minimum.

The situation changes in the presence of the magnetic field. Since $\tilde L\simeq-\beta r^2$ at large $r$, Eq.~(\ref{eq:R5}) gives $\mathcal{R}\simeq-\beta^2r^6$, and $\mathcal{R}\to-\infty$ for any energy. In the Newtonian limit with $a=0$ this corresponds to
\begin{equation}
U=\frac{L^2-\mu}{2r^2}-\beta L+\frac{\beta^2r^2}{2}\, ,
\end{equation}
where the last term grows at large distances and, for $L^2>\mu$, creates a potential well with the minimum at $r^4=(L^2-\mu)/\beta^2$. As a function of $x=r^2$, $\mathcal{R}$ is now a cubic polynomial, and a bound orbit corresponds to an interval $x_2<x<x_3$ outside the horizon where $\mathcal{R}>0$. The particle then moves between the periapsis $r_p=\sqrt{x_2}$ and the apoapsis $r_a=\sqrt{x_3}$. Thus, in $D=5$ dimensions, all bound orbits, and therefore all periodic orbits, exist only because of the magnetic field.
\begin{figure*}[t]
\centering
\includegraphics[width=\textwidth]{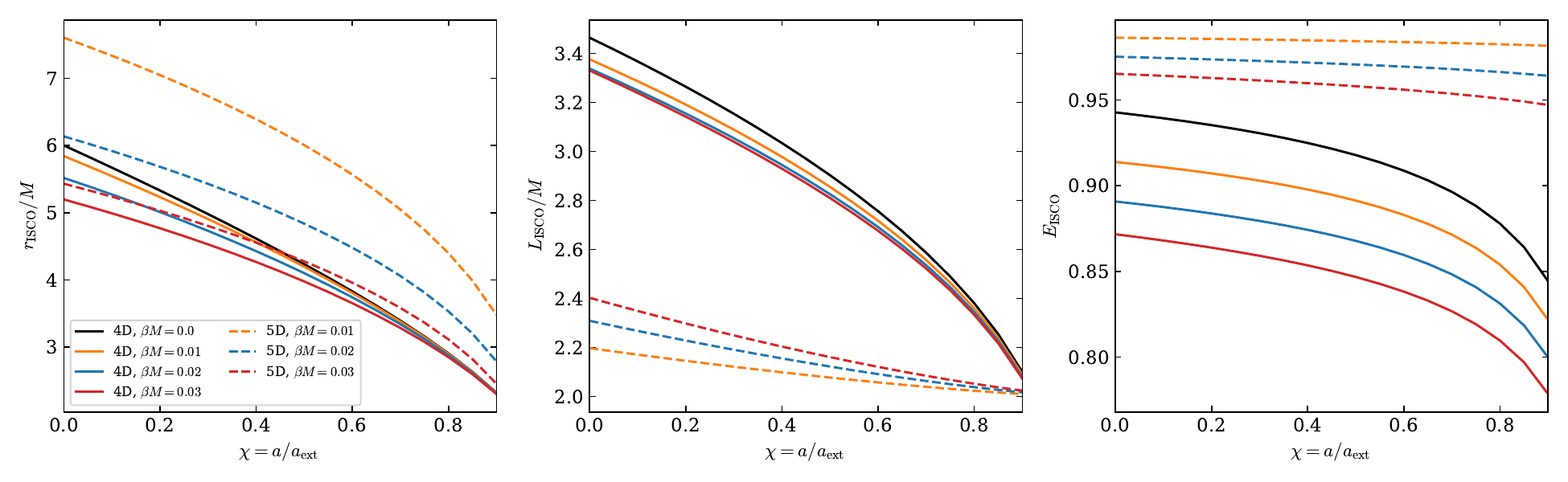}
\caption{Radius, angular momentum, and energy of the ISCO as functions of the spin parameter $\chi=a/a_{\rm ext}$. 
}
\label{fig:isco}
\end{figure*}
\begin{figure*}[t]
\centering
\includegraphics[width=\textwidth]{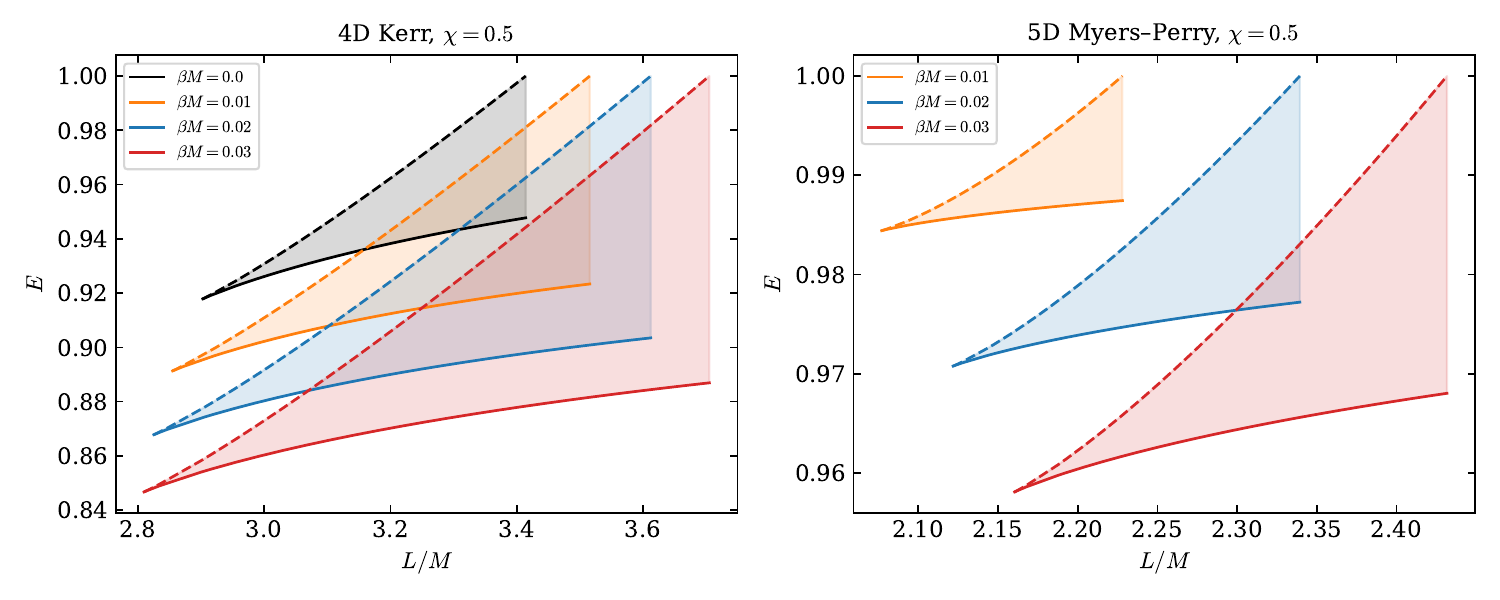}
\caption{
Region of bound orbits in the $(L,E)$ plane for $\chi=0.5$. 
Solid lines: stable circular orbits $V_{\min}(L)$; dashed lines: unstable circular orbits $V_{\max}(L)$. The range of $L$ is $L_{\rm ISCO}<L<L_*$.
}
\label{fig:EL}
\end{figure*}

\subsection{Circular orbits and the ISCO}\label{sec:circ}

Circular orbits correspond to the extrema of the effective potential $V_{\rm eff}$, and stable circular orbits correspond to its minima. The innermost stable circular orbit (ISCO) is determined by the conditions
\begin{equation}\label{eq:isco}
\frac{dV_{\rm eff}}{dr}=0\, ,\qquad \frac{d^2V_{\rm eff}}{dr^2}=0\, ,
\end{equation}
which we solve numerically to find $r_{\rm ISCO}$, $L_{\rm ISCO}$, and $E_{\rm ISCO}$. For $L>L_{\rm ISCO}$ the effective potential has a maximum $V_{\max}$ and a minimum $V_{\min}$, and bound orbits exist in the range
\begin{equation}\label{eq:boundrange}
V_{\min}<E<V_{\max}\, .
\end{equation}
The orbit with $E=V_{\max}$ is the separatrix: it asymptotically approaches the unstable circular orbit at the maximum of the potential. For $E>V_{\max}$ the particle falls into the BH. Since $V_{\rm eff}\simeq|\beta|\,r$ grows at large $r$, the upper limit of the bound energies is $V_{\max}$ rather than $E=1$.

In Fig.~\ref{fig:isco}, we show the ISCO parameters as functions of the spin parameter $\chi$. For the Kerr BH the magnetic field has a weak effect on the ISCO and moves it closer to the BH, while for the $D=5$ dimensional BH the ISCO exists only because of the magnetic field and is much more sensitive to it. For $\chi=0$, its radius increases from $5.4M$ to $7.6M$ as $\beta M$ decreases from $0.03$ to $0.01$, and it goes to infinity as $\beta\to0$. As $\beta$ increases, the angular momentum at the ISCO decreases for the Kerr BH and increases for the five-dimensional BH, as shown in Fig.~\ref{fig:isco}.
An important difference appears in the energy. The binding energy at the ISCO, $1-E_{\rm ISCO}$, is several times smaller for the five-dimensional BH: its ISCO energy is close to unity, $E_{\rm ISCO}\simeq0.95$--$0.99$, while for the Kerr BH $E_{\rm ISCO}\simeq0.78$--$0.94$ (Fig.~\ref{fig:isco}). This means that a charged particle is much more weakly bound to the five-dimensional BH. As the spin grows, the ISCO radii of the two BHs approach each other.

In Fig.~\ref{fig:EL}, we represent the region of bound orbits in the $(L,E)$ plane. It is bounded from below by the stable circular orbits, $E=V_{\min}(L)$, and from above by the unstable circular orbits, $E=V_{\max}(L)$, and the two boundaries meet at the ISCO. As $\beta$ increases, the region extends to larger $L$ and moves to smaller $E$ for both BHs. For the $D=5$ dimensional BH, the region is much narrower in $L$ and lies closer to $E=1$.

At the same $\beta$ and $\chi$, the depth of the potential well at $L=L_A$ for the five-dimensional BH is several times smaller than for the Kerr BH, as shown in Fig.~\ref{fig:EL}. Therefore, the periodic orbits of the $D=5$ dimensional BH that we wish to discuss in the next section lie much closer to the separatrix. For $\chi=0$ they are extremely close to it, and we present the periodic orbits for $\chi=0.5$. 
\begin{figure*}[t]
\centering
\includegraphics[width=\textwidth]{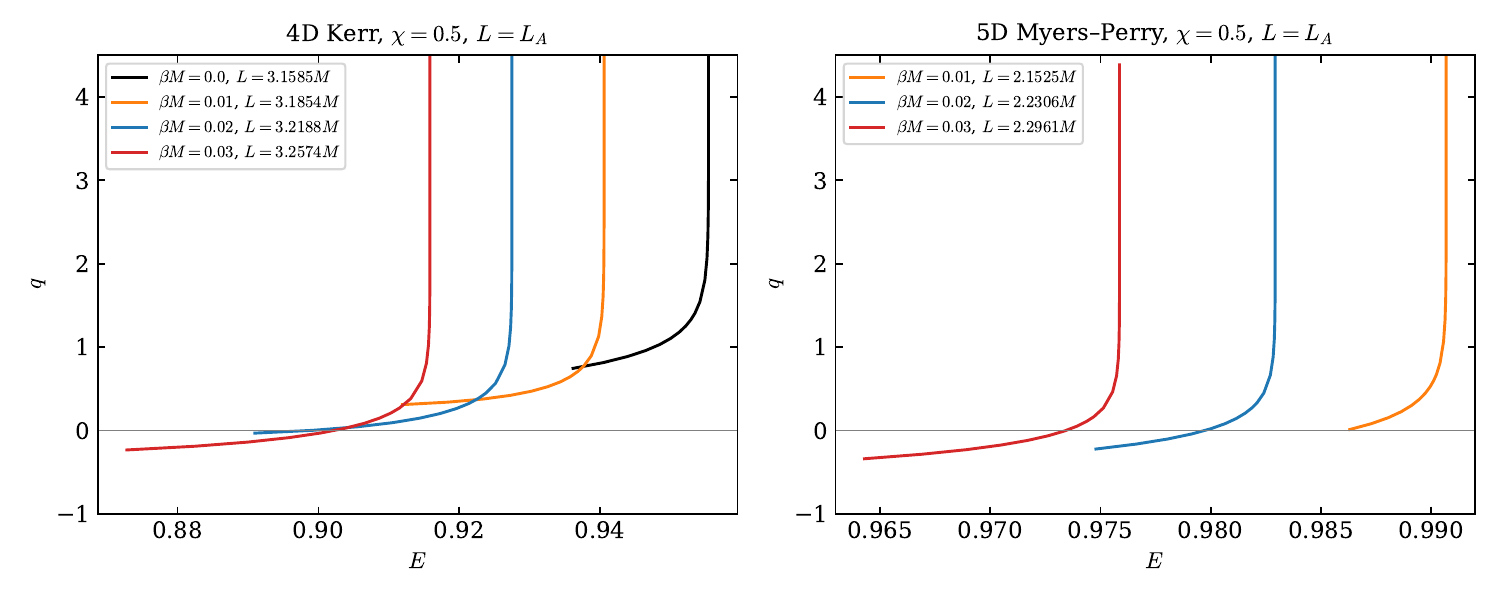}
\caption{Rational number $q$ as a function of the energy for the Kerr BH (left panel) and $D=5$ dimensional BH (right panel) in a uniform magnetic field. Here, we keep $L=L_A$ and $\chi=0.5$ fixed. }
\label{fig:qE}
\end{figure*}
\begin{figure*}[t]
\centering
\includegraphics[width=\textwidth]{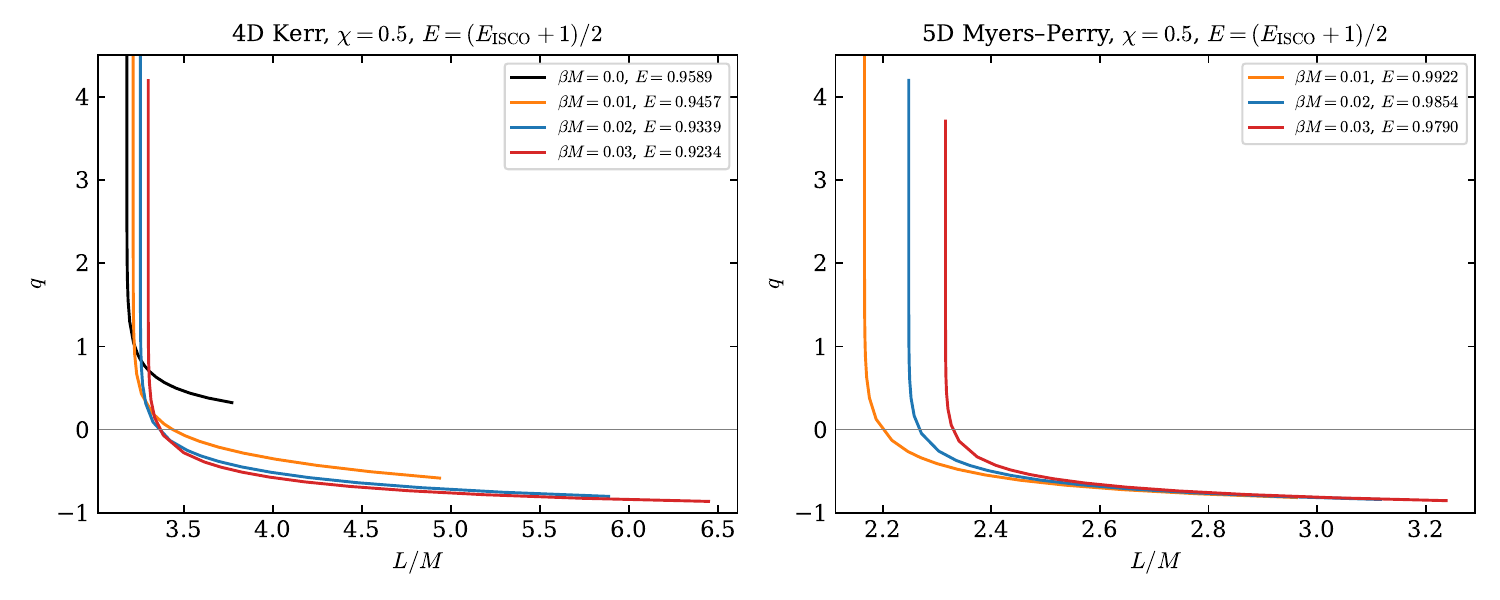}
\caption{Rational number $q$ as a function of the angular momentum for the Kerr BH (left panel) and $D=5$ dimensional BH (right panel) in a uniform magnetic field. Here, we keep $E=E_{\rm fix}=(E_{\rm ISCO}+1)/2$ and $\chi=0.5$ fixed.
}
\label{fig:qL}
\end{figure*}
\begin{table*}[htbp]
\centering
\resizebox{\textwidth}{!}{%
\begin{tabular}{|c|c|c|c|c|c|c|c|}
\hline
BH & $\beta M$ & $L_A/M$ & $V_{\max}$ & $E_{(1,1,0)}$ & $E_{(1,2,0)}$ & $E_{(2,1,1)}$ & $E_{(2,2,1)}$ \\ \hline
4D & 0.0 & 3.15853986 & 0.955423519040 & 0.947717680930 & 0.955150576392 & 0.954011259975 & 0.955369853365 \\ \hline
4D & 0.01 & 3.18537255 & 0.940585983455 & 0.939354545989 & 0.940551184841 & 0.940381916487 & 0.940580016213 \\ \hline
4D & 0.02 & 3.21875155 & 0.927486563311 & 0.927048858064 & 0.927476455587 & 0.927420266593 & 0.927485020393 \\ \hline
4D & 0.03 & 3.25741336 & 0.915844860098 & 0.915615674803 & 0.915840505715 & 0.915813308525 & 0.915844258901 \\ \hline
5D & 0.01 & 2.15252695 & 0.990679899453 & 0.990532878301 & 0.990674471545 & 0.990651754336 & 0.990678850582 \\ \hline
5D & 0.02 & 2.23061997 & 0.982924681847 & 0.982866621716 & 0.982923546543 & 0.982916571833 & 0.982924522853 \\ \hline
5D & 0.03 & 2.29607446 & 0.975862936742 & 0.975833169218 & 0.975862555990 & 0.975859571653 & 0.975862893655 \\ \hline
\hline
BH & $\beta M$ & $L_A/M$ & $V_{\max}$ & $E_{(3,1,2)}$ & $E_{(3,2,2)}$ & $E_{(4,1,3)}$ & $E_{(4,2,3)}$ \\ \hline
4D & 0.0 & 3.15853986 & 0.955423519040 & 0.954609750851 & 0.955392269799 & 0.954804868274 & 0.955399670247 \\ \hline
4D & 0.01 & 3.18537255 & 0.940585983455 & 0.940472967667 & 0.940582667191 & 0.940501826873 & 0.940583511177 \\ \hline
4D & 0.02 & 3.21875155 & 0.927486563311 & 0.927451156566 & 0.927485738666 & 0.927460684569 & 0.927485960431 \\ \hline
4D & 0.03 & 3.25741336 & 0.915844860098 & 0.915828556726 & 0.915844549364 & 0.915833140102 & 0.915844636702 \\ \hline
5D & 0.01 & 2.15252695 & 0.990679899453 & 0.990663645276 & 0.990679292978 & 0.990667544859 & 0.990679438281 \\ \hline
5D & 0.02 & 2.23061997 & 0.982924681847 & 0.982920471264 & 0.982924599278 & 0.982921647795 & 0.982924622345 \\ \hline
5D & 0.03 & 2.29607446 & 0.975862936742 & 0.975861309208 & 0.975862915901 & 0.975861804852 & 0.975862922248 \\ \hline
\end{tabular}}
\caption{Energies $E$ of the periodic orbits $(z,w,v)$ at fixed $L=L_A=(L_{\rm ISCO}+L_*)/2$ and $\chi=0.5$.
}
\label{tab:E05}
\end{table*}
\begin{table*}[htbp]
\centering
\resizebox{\textwidth}{!}{%
\begin{tabular}{|c|c|c|c|c|c|c|c|}
\hline
BH & $\beta M$ & $E$ & $L_{\rm sep}/M$ & $L_{(1,1,0)}/M$ & $L_{(1,2,0)}/M$ & $L_{(2,1,1)}/M$ & $L_{(2,2,1)}/M$ \\ \hline
4D & 0.0 & 0.95894100 & 3.179807848588 & 3.223375032946 & 3.181344964055 & 3.187858106529 & 3.180103700475 \\ \hline
4D & 0.01 & 0.94566852 & 3.215225349368 & 3.221149645723 & 3.215390336573 & 3.216212961281 & 3.215252924419 \\ \hline
4D & 0.02 & 0.93390643 & 3.255571698536 & 3.257761175793 & 3.255620103196 & 3.255897182954 & 3.255578897604 \\ \hline
4D & 0.03 & 0.92336220 & 3.299662735764 & 3.300819417278 & 3.299683724966 & 3.299818537886 & 3.299665563431 \\ \hline
5D & 0.01 & 0.99220858 & 2.166445915549 & 2.167560552396 & 2.166483195090 & 2.166649941048 & 2.166452724220 \\ \hline
5D & 0.02 & 0.98538424 & 2.247936333074 & 2.248283707536 & 2.247942495252 & 2.247982609153 & 2.247937153574 \\ \hline
5D & 0.03 & 0.97904321 & 2.315632198075 & 2.315790033449 & 2.315634044204 & 2.315649269265 & 2.315632397717 \\ \hline
\hline
BH & $\beta M$ & $E$ & $L_{\rm sep}/M$ & $L_{(3,1,2)}/M$ & $L_{(3,2,2)}/M$ & $L_{(4,1,3)}/M$ & $L_{(4,2,3)}/M$ \\ \hline
4D & 0.0 & 0.95894100 & 3.179807848588 & 3.184435492257 & 3.179978757464 & 3.183319240599 & 3.179937754663 \\ \hline
4D & 0.01 & 0.94566852 & 3.215225349368 & 3.215769219465 & 3.215240539110 & 3.215628967790 & 3.215236623117 \\ \hline
4D & 0.02 & 0.93390643 & 3.255571698536 & 3.255744139939 & 3.255575512773 & 3.255697215044 & 3.255574474882 \\ \hline
4D & 0.03 & 0.92336220 & 3.299662735764 & 3.299742604790 & 3.299664185335 & 3.299719920762 & 3.299663773639 \\ \hline
5D & 0.01 & 0.99220858 & 2.166445915549 & 2.166561701307 & 2.166449778449 & 2.166533136579 & 2.166448825180 \\ \hline
5D & 0.02 & 0.98538424 & 2.247936333074 & 2.247959964534 & 2.247936752053 & 2.247953220119 & 2.247936632472 \\ \hline
5D & 0.03 & 0.97904321 & 2.315632198075 & 2.315640331353 & 2.315632293190 & 2.315637811984 & 2.315632263726 \\ \hline
\end{tabular}}
\caption{Angular momenta $L$ of the periodic orbits $(z,w,v)$ at fixed $E=(E_{\rm ISCO}+1)/2$ and $\chi=0.5$.
}
\label{tab:L05}
\end{table*}

\section{Periodic orbits}\label{sec:periodic}

\textit{Units and parameters:} For further analysis of periodic orbits, we shall compare the two BHs to measure lengths in units of $M$. For example, for the $D=5$ dimensional BH, we set $\mu=4M^2$, so that the horizon radius of the nonrotating BH is $r_+=2M$ in both cases. This condition fixes only the length scale, i.e. it does not mean that the two BHs have the same physical mass, since the $D=5$ dimensional mass $M_5=3\pi\mu/(8G_5)$ involves the five-dimensional Newton constant. For the spin, we use the dimensionless parameter $\chi=a/a_{\rm ext}$, with $a_{\rm ext}=M$ for Kerr and $a_{\rm ext}=\sqrt\mu=2M$ for the $D=5$ dimensional BH, and we compare the two BHs at the same $\chi$. The magnetic parameter $\beta$ is given in units of $1/M$; we use $\beta M=0$, $0.01$, $0.02$, and $0.03$, where $\beta=0$ is considered only for Kerr. 

It must be emphasized that a bound orbit is characterized by two frequencies, the radial frequency $\omega_r$ and the azimuthal frequency $\omega_\phi$, and it is periodic when their ratio is a rational number. Following Ref.~\cite{Levin_2008}, we write this ratio as
\begin{equation}\label{eq:qdef}
q=\frac{\omega_\phi}{\omega_r}-1=w+\frac{v}{z}\, ,
\end{equation}
where $z$, $w$, and $v$ are the zoom, whirl, and vertex numbers. During one radial period the angle $\phi$ changes by $2\pi(1+q)$. Since the motion from the periapsis $r_p$ to the apoapsis $r_a$ takes half of the radial period, $q$ can be found from
\begin{equation}\label{eq:qint}
q=\frac{1}{\pi}\int_{r_p}^{r_a}\frac{d\phi}{dr}\,dr-1\, ,\qquad \frac{d\phi}{dr}=\frac{\dot\phi}{\dot r}=\frac{r^2\dot\phi}{\sqrt{\mathcal{R}}}\, ,
\end{equation}
where $\dot\phi$ is given by Eq.~(\ref{eq:dot5}) for the five-dimensional BH.

When the energy approaches $V_{\min}$, the orbit becomes nearly circular, and $q$ tends to the value $q_0$ determined by the angular velocity of the stable circular orbit and the frequency $\omega_r$ of small radial oscillations around it,
\begin{equation}\label{eq:q0}
q_0=\frac{\dot\phi(r_{\min})}{\omega_r}-1\, ,\qquad \omega_r^2=-\frac{1}{2}\,\frac{d^2\Psi}{dr^2}\bigg|_{r=r_{\min}},
\end{equation}
where $r_{\min}$ is the position of the minimum of $V_{\rm eff}$, and $\Psi$ is evaluated at $E=V_{\min}$. When the energy approaches $V_{\max}$, the particle spends more time near the unstable circular orbit, and $q$ grows without limit. Therefore, orbits with large $w$ have energies very close to $V_{\max}$. In the presence of the magnetic field, $q_0$ can be negative, which means $\omega_\phi<\omega_r$. In our analysis, we consider only orbits with $q>0$.

To find the periodic orbits, we fix either the angular momentum or the energy. In the first case, we fix the angular momentum at
\begin{equation}\label{eq:LA}
L_A=\frac{1}{2}\left(L_{\rm ISCO}+L_*\right),
\end{equation}
where $L_*$ is the angular momentum for which $V_{\max}=1$~\cite{Alloqulov25GW,Xamidov2026SBR}. For the Kerr BH without magnetic field, $L_*$ is the angular momentum of the marginally bound orbit. For $\beta\neq0$ this definition is a convenient convention, which we use for both BHs. In the second case, we fix the energy at
\begin{equation}\label{eq:Efix}
E_{\rm fix}=\frac{1}{2}\left(E_{\rm ISCO}+1\right).
\end{equation}
For $E=E_{\rm fix}$ bound orbits exist for $L_{\rm sep}<L<L_{\rm top}$, where $L_{\rm sep}$ and $L_{\rm top}$ are defined by $V_{\max}(L_{\rm sep})=E_{\rm fix}$ and $V_{\min}(L_{\rm top})=E_{\rm fix}$. As $L\to L_{\rm sep}$, the orbit approaches the separatrix and $q$ grows without limit.
\begin{figure*}[!t]
\centering
\includegraphics[width=0.95\textwidth]{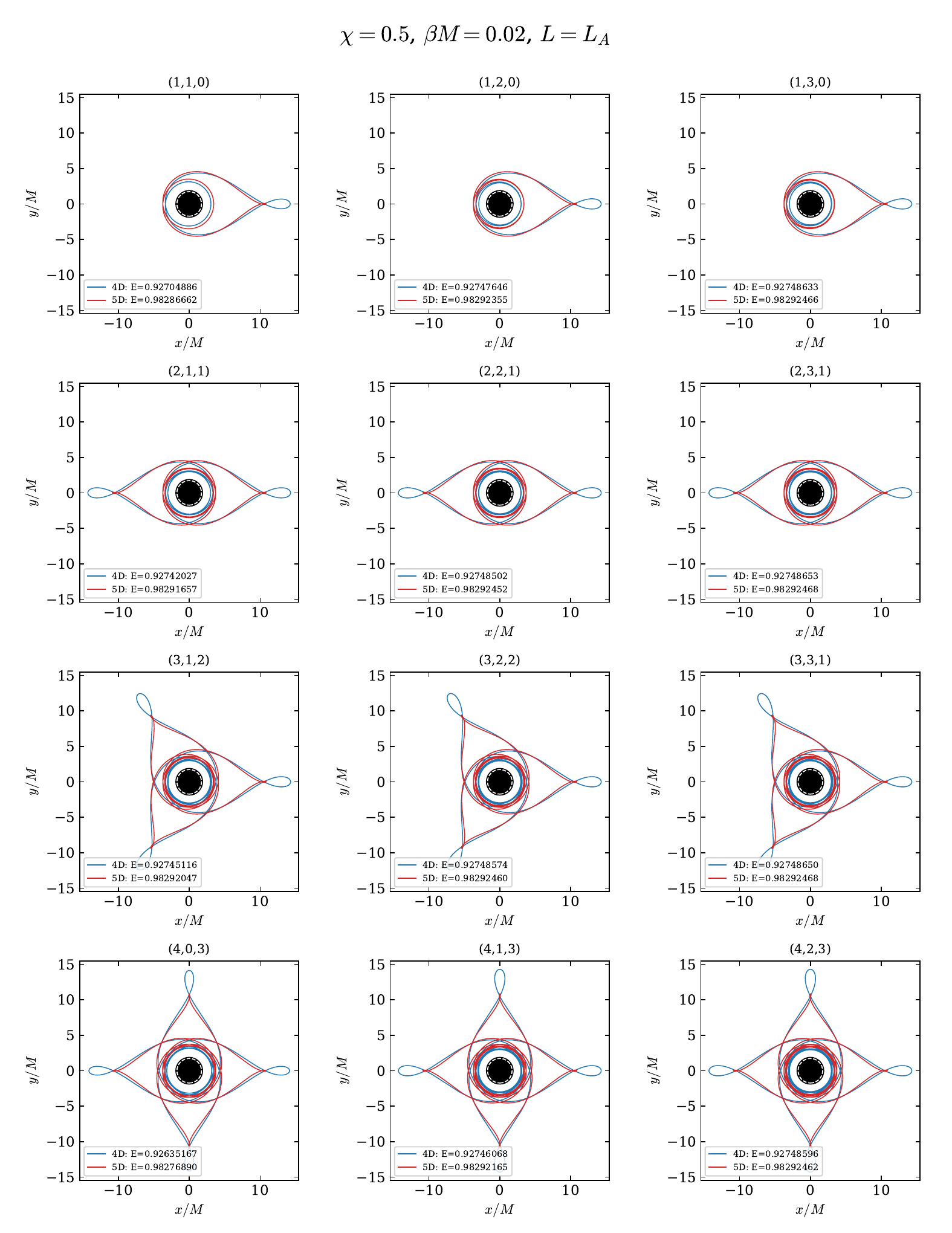}
\caption{Periodic orbits $(z,w,v)$ for $\chi=0.5$, $\beta M=0.02$, and fixed $L=L_A$. The black disk and the white dashed circle show the horizons of the Kerr and five-dimensional BHs.}
\label{fig:orbL05}
\end{figure*}

\begin{figure*}[!t]
\centering
\includegraphics[width=0.95\textwidth]{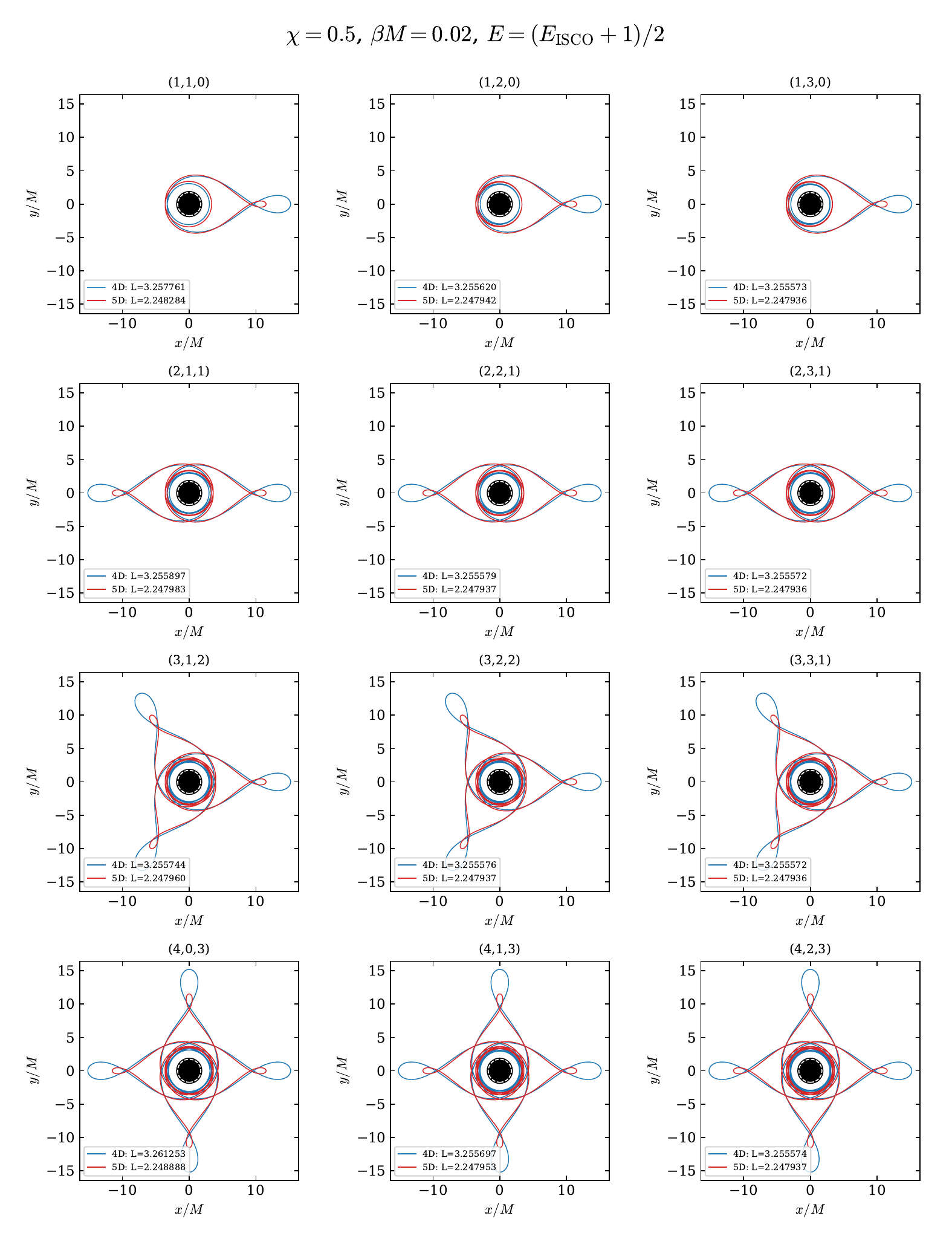}
\caption{
Periodic orbits $(z,w,v)$ for $\chi=0.5$, $\beta M=0.02$, and fixed $E=E_{\rm fix}$. The black disk and the white dashed circle show the horizons of the Kerr and five-dimensional BHs.
}
\label{fig:orbE05}
\end{figure*}

In Fig.~\ref{fig:qE}, we show $q$ as a function of the energy at $L=L_A$. As can be seen in Fig.~\ref{fig:qE}, each curve starts at $q_0$ at the bottom of the potential well and grows without limit as $E\to V_{\max}$. The magnetic field shifts the curves to smaller energies and lowers $q_0$. For the $D=5$ dimensional BH, the increase of $q$ occurs in a very narrow energy interval near $V_{\max}$, so that the orbits with $w\ge1$ lie much closer to the separatrix than for the Kerr BH. In addition, in Fig.~\ref{fig:qL} we show $q$ as a function of the angular momentum at $E=E_{\rm fix}$. $q$ grows without limit as $L\to L_{\rm sep}$ and decreases as $L$ increases. In the presence of the magnetic field, $q$ becomes negative at large $L$, i.e., $\omega_\phi<\omega_r$. For the five-dimensional BH it reaches about $-0.8$ near $L_{\rm top}$, which reflects the Larmor motion of the charge in the magnetic field. For the Kerr BH without magnetic field $q$ remains positive.

We also list the periodic orbits with eight sets $(z,w,v)$ for $\chi=0.5$ in Tables~\ref{tab:E05} and \ref{tab:L05}. We provide their energies at fixed $L=L_A$ in Table~\ref{tab:E05}, while we give their angular momenta at fixed $E=E_{\rm fix}$ in Table~\ref{tab:L05}. For the $D=5$ dimensional BH the orbits with $w=2$ lie extremely close to the separatrix, i.e. their energies are only $10^{-8}$--$10^{-5}$ below $V_{\max}$. For this reason, the periodic orbits must be found with high numerical precision.

In Figs.~\ref{fig:orbL05} and \ref{fig:orbE05}, we show the periodic orbits for $\chi=0.5$ and $\beta M=0.02$ at fixed $L=L_A$ and fixed $E=E_{\rm fix}$, respectively. Each panel contains the orbits around the Kerr BH (blue) and around the $D=5$ dimensional BH (red) with the same $(z,w,v)$. As usual, larger $z$ gives more leaves, and larger $w$ gives more revolutions near the BH. Near the BH, in the whirl part, the orbits of the two BHs are very similar. Far from the BH the orbits differ, i.e. the orbits around the Kerr BH make loops near the apoapsis, while the orbits around the $D=5$ dimensional rotating BH form only very small loops. These loops are caused by the magnetic field. According to Eq.~(\ref{eq:dot5}), $\dot\phi$ changes sign close to the radius where $\tilde L=0$, i.e., at $r^2\simeq L/\beta-a^2$. A loop forms when the apoapsis lies beyond this radius, and for the $D=5$ dimensional BH the apoapsis lies only slightly beyond it, and the loops are small. Such curly orbits are typical for charged particles around magnetized BHs~\cite{FrolovShoom2010}.

\section{Gravitational waves}\label{sec:gw}

We consider an EMRI in which a compact object of mass $m$ moves along a periodic orbit around a supermassive BH of mass $M\gg m$. We first neglect the radiation reaction and treat the orbit as a solution of the equations of motion, and the time during which this approximation is valid is estimated below. The waveforms are computed with the numerical kludge method~\cite{Babak2007Kludge,Yang24JCAP,2025EPJC...85...36Z}. The particle trajectory is obtained by solving Eqs.~(\ref{eq:dot5}) and (\ref{eq:rddot}). The Boyer--Lindquist coordinates are then mapped to flat Cartesian coordinates, $x=r\cos\phi$, $y=r\sin\phi$, $z=0$, and the corresponding gravitational waveform is computed using the quadrupole formula
\begin{equation}\label{eq:hij}
h_{ij}=\frac{2G}{c^4D_L}\,\frac{d^2I_{ij}}{dt^2}\, ,\qquad I_{ij}=m\,x_ix_j\, ,
\end{equation}
where $D_L$ is the luminosity distance and $t$ is the coordinate time, i.e., the time of a distant observer.

We project $h_{ij}$ onto the detector frame~\cite{Poisson_Will_2014,Yang24JCAP}
\begin{eqnarray}
e_X&=&[\cos\zeta,-\sin\zeta,0]\, ,\nonumber\\
e_Y&=&[\cos\iota\sin\zeta,\cos\iota\cos\zeta,-\sin\iota]\, ,
\end{eqnarray}
where $\iota$ is the inclination and $\zeta$ is the longitude of the pericenter. The polarizations $h_+=\frac{1}{2}(e_X^ie_X^j-e_Y^ie_Y^j)h_{ij}$ and $h_\times=\frac{1}{2}(e_X^ie_Y^j+e_Y^ie_X^j)h_{ij}$ then take the form 
\begin{equation}\label{eq:hplus}
h_{+,\times}=\frac{Gm}{2c^4D_L}\,\frac{d^2H_{+,\times}}{dt^2}\, ,
\end{equation}
where $\Phi=\phi+\zeta$ and
\begin{eqnarray}\label{eq:hcross}
H_+&=&r^2\left[\sin^2\iota+(1+\cos^2\iota)\cos2\Phi\right],\nonumber\\
H_\times&=&2r^2\cos\iota\,\sin2\Phi\, .
\end{eqnarray}
The trace part of $I_{ij}$ does not contribute to $h_+$ and $h_\times$.

For simplicity, we denote $r^2$ by $P$, so that $P'=2rr'$ and $P''=2(r'^2+rr'')$, where the primes denote derivatives with respect to $t$. Then we obtain
\predisplaypenalty=0
\begin{eqnarray}
h_+&=&\frac{Gm}{2c^4D_L}\Big\{\sin^2\iota\,P''\nonumber\\
&&+(1+\cos^2\iota)\Big[\left(P''-4P\Phi'^2\right)\cos2\Phi\nonumber\\
&&-\left(4P'\Phi'+2P\Phi''\right)\sin2\Phi\Big]\Big\},\label{eq:hpexp}
\end{eqnarray}
\predisplaypenalty=10000
\begin{eqnarray}
h_\times&=&\frac{Gm\cos\iota}{c^4D_L}\Big[\left(P''-4P\Phi'^2\right)\sin2\Phi\nonumber\\
&&+\left(4P'\Phi'+2P\Phi''\right)\cos2\Phi\Big].\label{eq:hxexp}
\end{eqnarray}

Here $r''$ and $\phi''$ are taken from the equations of motion, so that the acceleration includes both the gravitational and the magnetic forces. The derivatives with respect to $t$ are obtained from those with respect to $\tau$ using $d/dt=\dot t^{-1}d/d\tau$. 
\begin{figure*}[!t]
\centering
\includegraphics[width=0.95\textwidth]{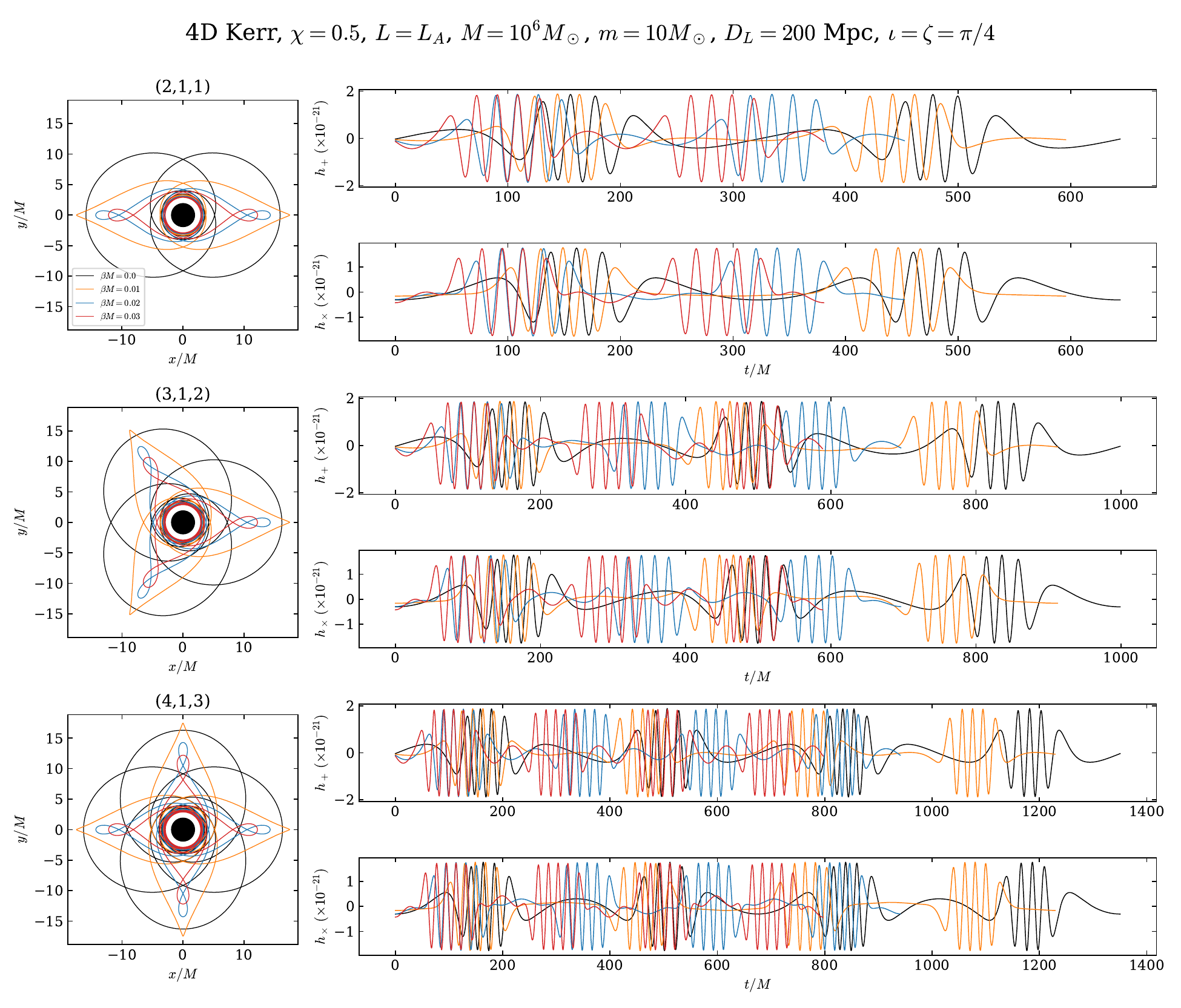}
\caption{Periodic orbits (left) and the corresponding polarizations $h_+$ and $h_\times$ (right) for the Kerr BH with $\chi=0.5$ and $L=L_A$. 
$M=10^6M_\odot$, $m=10M_\odot$, $D_L=200$\,Mpc, $\iota=\zeta=\pi/4$; one full period is shown.}
\label{fig:wf4D}
\end{figure*}
\begin{figure*}[!t]
\centering
\includegraphics[width=0.95\textwidth]{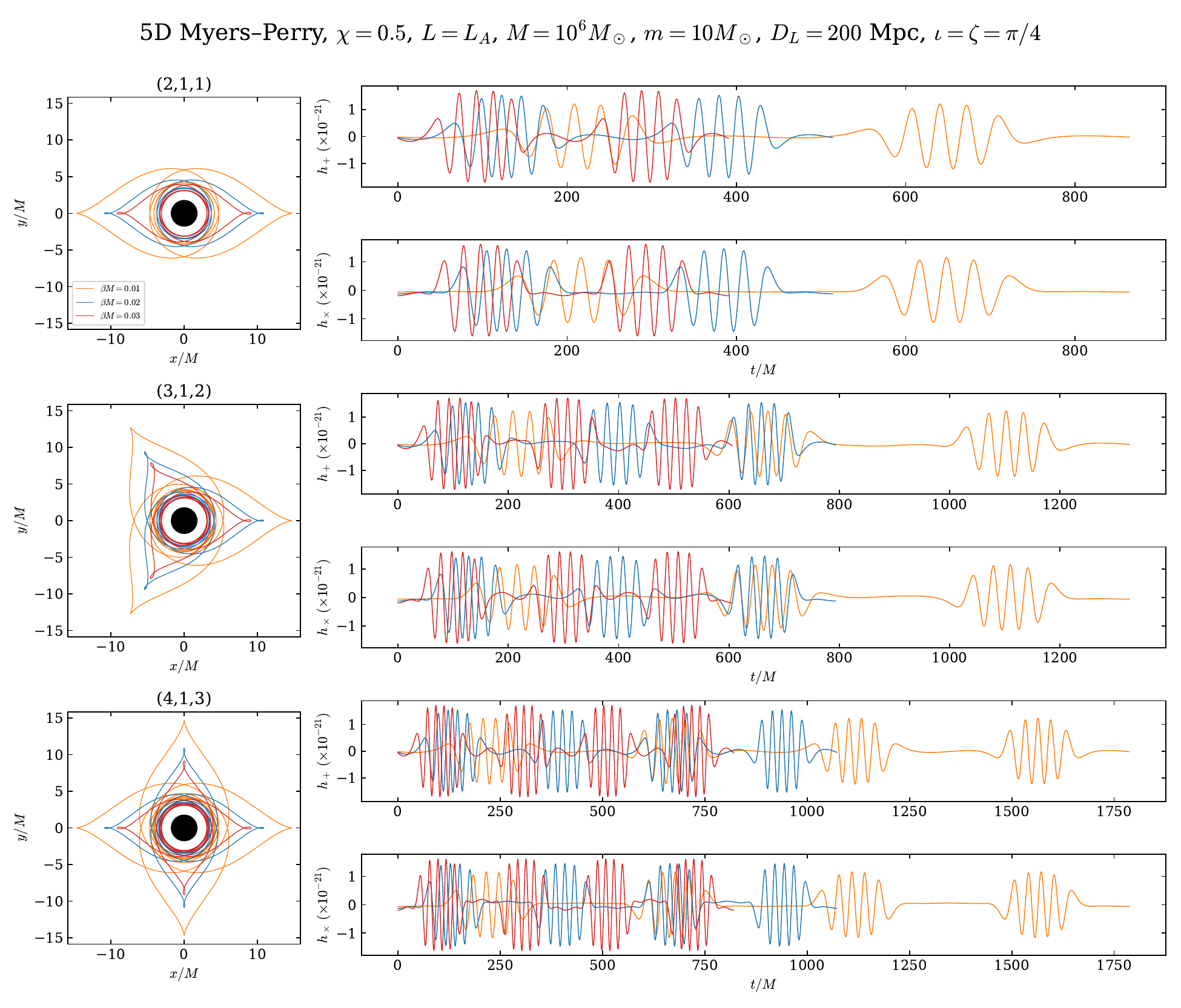}
\caption{
Periodic orbits (left) and the corresponding polarizations $h_+$ and $h_\times$ (right) five-dimensional BH with $\chi=0.5$ and $L=L_A$. 
$M=10^6M_\odot$, $m=10M_\odot$, $D_L=200$\,Mpc, $\iota=\zeta=\pi/4$; one full period is shown.
}
\label{fig:wf5D}
\end{figure*}
\begin{figure*}[t]
\centering
\includegraphics[width=0.9\textwidth]{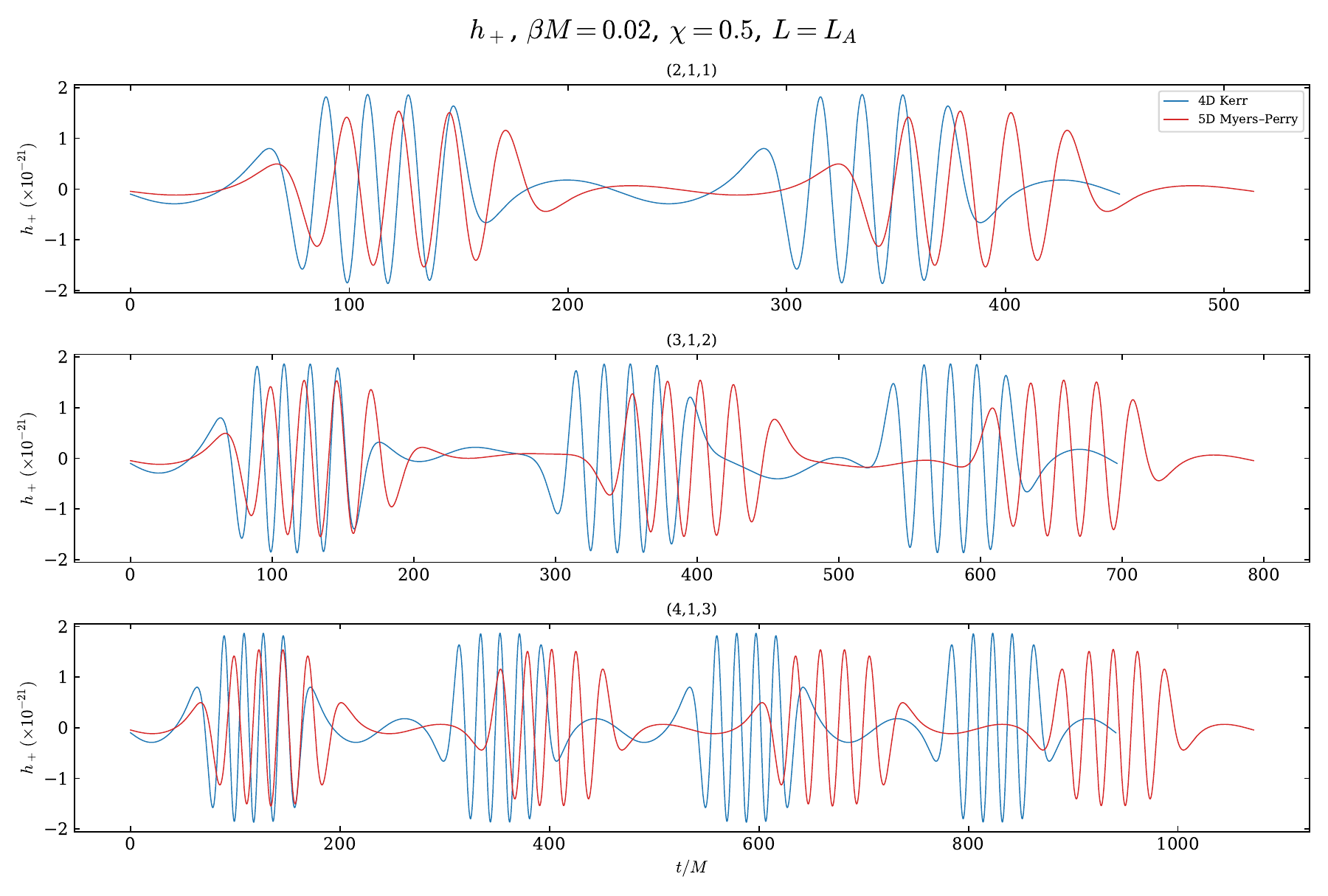}
\caption{Polarization $h_+$ of the $(2,1,1)$, $(3,1,2)$, and $(4,1,3)$ orbits for the Kerr BH (blue) and the five-dimensional BH (red) at $\beta M=0.02$, $\chi=0.5$, and $L=L_A$.}
\label{fig:wfcomp}
\end{figure*}

In units where $r$ is measured in $GM/c^2$ and $t$ in $GM/c^3$. In Eq.~(\ref{eq:hplus}), we keep its form with the factor $Gm/(c^4D_L)$ replaced by $Gm/(c^2D_L)$. We take $M=10^6M_\odot$, $m=10M_\odot$, $D_L=200$\,Mpc, and $\iota=\zeta=\pi/4$, for which $Gm/(c^2D_L)=2.39\times10^{-21}$ and $GM/c^3=4.93$\,s. 
\cred{For the $D=5$ dimensional BH we use the same four-dimensional quadrupole formula, and this is the main approximation of our GW analysis. In a $D=5$ dimensional asymptotically flat spacetime GWs have five polarizations, their amplitude decays as $r^{-3/2}$ instead of $r^{-1}$, and the radiation does not propagate only along the light cone~\cite{CardosoDiasLemos2003}. A consistent treatment would therefore require the $D=5$ dimensional wave zone, or a specific model of how the waves reach a four-dimensional observer. Instead, we use the five-dimensional metric only to describe the orbital motion and compute the waveforms of both BHs with the same formula. In this approach, the properties determined by the orbital motion, such as the period, the frequencies of the spectral lines, and the zoom--whirl structure of the signal, can be compared directly between the two BHs. The amplitude, the signal-to-noise ratio, and the detectability depend on the treatment of the wave zone, and for the five-dimensional BH they should be regarded as illustrative.
}
In Figs.~\ref{fig:wf4D} and \ref{fig:wf5D}, we show the waveforms of the $(2,1,1)$, $(3,1,2)$, and $(4,1,3)$ orbits for $\chi=0.5$ and $L=L_A$, while we summarize their main properties in Table~\ref{tab:gw}. The zoom and whirl phases are clearly seen in the waveforms. The slowly varying parts of the signal are emitted near the apoapsis, while the bursts of fast oscillations with the largest amplitude are emitted during the whirls near the BH.

For both the Kerr and $D=5$ dimensional BHs in the presence of an extrenal magnetic field, as $\beta$ increases, the orbital period decreases, the periapsis moves closer to the BH, and the strongest spectral lines move to higher frequencies. These changes are larger for the $D=5$ dimensional BH. For example, for the $(2,1,1)$ orbit, the frequency of the strongest line increases from $8.2$ to $10.7$\,mHz for the Kerr BH as $\beta M$ changes from $0$ to $0.03$, and from $5.6$ to $9.4$\,mHz for the $D=5$ dimensional BH as $\beta M$ changes from $0.01$ to $0.03$, as shown in Table~\ref{tab:gw}.

In Fig.~\ref{fig:wfcomp}, we compare the waveforms of two BHs for $\beta M=0.02$. Compared to the Kerr signal, the $D=5$ dimensional signal has a longer period, lower frequencies, and an amplitude reduced by about $18\%$, because its whirl phase occurs at a larger distance from the BH. As the magnetic field parameter increases, the differences between the waveforms become smaller, in agreement with the behavior of the ISCO.
\begin{table*}[t]
\caption{
Properties of the GW signals for $\chi=0.5$ and $L=L_A$. The signal-to-noise ratios are computed over the effective observation time $T_{\rm eff}=N_{\rm eff}T$; the last column gives, for comparison, the LISA value for $T_{\rm eff}=1$ week. The time unit is $GM/c^3=4.93$\,s for $M=10^6M_\odot$.
}
\label{tab:gw}
\begin{ruledtabular}
\begin{tabular}{ccccccccccccc}
BH & $\beta M$ & $(z,w,v)$ & $T/M$ & $r_p/M$ & $r_a/M$ & $f_{\rm peak}$ [mHz] & $h_{\max}/10^{-21}$ & $N_{\rm eff}$ & $\rho_{\rm LISA}$ & $\rho_{\rm Taiji}$ & $\rho_{\rm TianQin}$ & $\rho_{\rm LISA}^{\rm 1\,week}$ \\ \hline
Kerr & 0.0 & (2,1,1) & 643.9 & 3.450 & 15.886 & 8.20 & 1.859 & 190 & 79.4 & 161.3 & 65.5 & 79.4 \\
Kerr & 0.0 & (3,1,2) & 998.7 & 3.367 & 16.206 & 8.13 & 1.883 & 122 & 81.4 & 166.1 & 69.2 & 81.4 \\
Kerr & 0.0 & (4,1,3) & 1350.5 & 3.336 & 16.311 & 8.72 & 1.890 & 89 & 81.9 & 167.5 & 70.4 & 82.4 \\
\hline
Kerr & 0.01 & (2,1,1) & 595.5 & 3.127 & 17.433 & 9.55 & 1.879 & 81 & 53.9 & 111.7 & 50.0 & 85.9 \\
Kerr & 0.01 & (3,1,2) & 912.4 & 3.103 & 17.461 & 9.57 & 1.885 & 30 & 41.7 & 86.6 & 39.1 & 88.2 \\
Kerr & 0.01 & (4,1,3) & 1229.1 & 3.094 & 17.469 & 9.58 & 1.887 & 16 & 35.8 & 74.4 & 33.8 & 89.1 \\
Kerr & 0.02 & (2,1,1) & 452.2 & 2.994 & 14.291 & 9.88 & 1.865 & 32 & 33.9 & 70.6 & 32.8 & 98.6 \\
Kerr & 0.02 & (3,1,2) & 696.5 & 2.982 & 14.295 & 9.91 & 1.868 & 11 & 25.2 & 52.6 & 24.6 & 100.8 \\
Kerr & 0.02 & (4,1,3) & 940.7 & 2.977 & 14.296 & 9.93 & 1.868 & 6 & 21.9 & 45.6 & 21.4 & 101.8 \\
Kerr & 0.03 & (2,1,1) & 380.6 & 2.903 & 12.203 & 10.67 & 1.849 & 19 & 25.9 & 54.0 & 25.7 & 106.5 \\
Kerr & 0.03 & (3,1,2) & 588.7 & 2.895 & 12.205 & 10.69 & 1.850 & 6 & 18.4 & 38.5 & 18.4 & 108.5 \\
Kerr & 0.03 & (4,1,3) & 796.8 & 2.892 & 12.205 & 10.70 & 1.851 & 3 & 15.3 & 31.9 & 15.3 & 109.7 \\
\hline
5D & 0.01 & (2,1,1) & 864.3 & 4.073 & 14.661 & 5.64 & 1.211 & 49 & 31.2 & 61.6 & 19.7 & 53.2 \\
5D & 0.01 & (3,1,2) & 1325.8 & 4.045 & 14.670 & 6.13 & 1.230 & 18 & 24.4 & 48.4 & 15.7 & 55.2 \\
5D & 0.01 & (4,1,3) & 1787.4 & 4.034 & 14.673 & 6.13 & 1.236 & 10 & 21.5 & 42.7 & 14.0 & 56.1 \\
5D & 0.02 & (2,1,1) & 513.5 & 3.397 & 10.851 & 7.91 & 1.536 & 9 & 15.7 & 32.5 & 13.5 & 80.9 \\
5D & 0.02 & (3,1,2) & 792.9 & 3.388 & 10.852 & 7.94 & 1.546 & 3 & 11.6 & 24.0 & 10.0 & 82.9 \\
5D & 0.02 & (4,1,3) & 1072.2 & 3.384 & 10.852 & 7.95 & 1.549 & 1 & 7.9 & 16.4 & 6.8 & 84.1 \\
5D & 0.03 & (2,1,1) & 390.5 & 3.086 & 9.122 & 9.36 & 1.708 & 3 & 9.5 & 19.9 & 9.1 & 97.2 \\
5D & 0.03 & (3,1,2) & 605.5 & 3.082 & 9.122 & 9.39 & 1.714 & 1 & 7.0 & 14.6 & 6.7 & 99.4 \\
5D & 0.03 & (4,1,3) & 820.4 & 3.080 & 9.123 & 9.40 & 1.716 & 1 & 8.2 & 17.2 & 7.9 & 100.4 \\
\end{tabular}
\end{ruledtabular}
\end{table*}

{\bf\textit{Radiation reaction and observation time:}} The periodic orbits are solutions of the conservative equations of motion, while the emission of GWs changes $E$ and $L$. To estimate how long a periodic orbit survives, we compute the energy and angular momentum carried away during one period. In the quadrupole approximation, the corresponding fluxes are defined by  
\begin{equation}\label{eq:fluxes}
F_E=\frac{1}{5}\left\langle Q'''_{ij}Q'''_{ij}\right\rangle,\qquad
F_L=\frac{2}{5}\,\epsilon_{zjk}\left\langle Q''_{j\ell}Q'''_{k\ell}\right\rangle\, ,
\end{equation}
where $Q_{ij}=m(x_ix_j-\delta_{ij}r^2/3)$ is evaluated along the same trajectory as the waveform, the primes denote derivatives with respect to $t$, and the brackets denote the average over one period (units are set to $G=c=1$). We denote by $\Delta E$ and $\Delta L$ the energy and angular momentum per unit mass lost during one period.

Here, the distance of an orbit to the separatrix is $\delta=V_{\max}(L)-E$. Since $dV_{\max}/dL=\Omega_u$ ($\partial V_{\rm eff}/\partial r=0$ at the maximum and $\partial V_{\rm eff}/\partial L=\dot\phi/\dot t$ at a turning point), where $\Omega_u$ is the angular velocity of the unstable circular orbit at the maximum of $V_{\rm eff}$, the change of $\delta$ during one period is given by 
\begin{equation}\label{eq:ddelta}
\Delta\delta=\Delta E-\Omega_u\,\Delta L\, .
\end{equation}
For all orbits considered here, we find $\Delta\delta<0$, i.e., the radiation reaction drives the orbit toward the separatrix. The number of periods before the separatrix is reached is then about $N_{\rm sep}=\delta/|\Delta\delta|$. This estimate assumes that $\Delta\delta$ is the same in every period, which is a good approximation when the orbit changes little during one period. When $|\Delta\delta|/\delta$ exceeds about $10\%$, the orbit changes noticeably already during one period, and $N_{\rm sep}$ gives only the order of magnitude. We therefore take the effective observation time of each orbit as $T_{\rm eff}=N_{\rm eff}T$, where $N_{\rm eff}=\min(\lfloor T_{\rm obs}/T\rfloor,\lfloor N_{\rm sep}\rfloor)$, but not less than one, and $T_{\rm obs}=1$ week. During this time we assume that the orbit remains periodic.
\begin{table*}[t]
\caption{Radiation reaction estimate for $\chi=0.5$ and $L=L_A$ ($m=10M_\odot$, $M=10^6M_\odot$): distance to the separatrix $\delta=V_{\max}-E$, its relative change during one period, Eq.~(\ref{eq:ddelta}), the number of periods $N_{\rm sep}$ and the time $t_{\rm sep}$ before the separatrix is reached, and the number of periods in one week. $^*$: $|\Delta\delta|/\delta>10\%$, the first-order estimate is rough; $^\dagger$: $N_{\rm sep}<1$.}
\label{tab:rad}
\begin{ruledtabular}
\begin{tabular}{cccccccc}
BH & $\beta M$ & $(z,w,v)$ & $\delta$ & $|\Delta\delta|/\delta$ [\%] & $N_{\rm sep}$ & $t_{\rm sep}$ [h] & periods in 1 week \\ \hline
Kerr & 0.0 & (2,1,1) & $1.41\times10^{-3}$ & 0.24 & 411.2 & 362.3 & 190 \\
Kerr & 0.0 & (3,1,2) & $8.14\times10^{-4}$ & 0.63 & 157.8 & 215.6 & 122 \\
Kerr & 0.0 & (4,1,3) & $6.19\times10^{-4}$ & 1.11 & 89.9 & 166.2 & 90 \\
\hline
Kerr & 0.01 & (2,1,1) & $2.04\times10^{-4}$ & 1.23 & 81.3 & 66.3 & 206 \\
Kerr & 0.01 & (3,1,2) & $1.13\times10^{-4}$ & 3.33 & 30.0 & 37.5 & 134 \\
Kerr & 0.01 & (4,1,3) & $8.42\times10^{-5}$ & 5.97 & 16.8 & 28.2 & 99 \\
Kerr & 0.02 & (2,1,1) & $6.63\times10^{-5}$ & 3.05 & 32.8 & 20.3 & 271 \\
Kerr & 0.02 & (3,1,2) & $3.54\times10^{-5}$ & 8.56 & 11.7 & 11.1 & 176 \\
Kerr & 0.02 & (4,1,3) & $2.59\times10^{-5}$ & 15.61$^*$ & 6.4 & 8.2 & 130 \\
Kerr & 0.03 & (2,1,1) & $3.16\times10^{-5}$ & 5.18 & 19.3 & 10.1 & 322 \\
Kerr & 0.03 & (3,1,2) & $1.63\times10^{-5}$ & 15.04$^*$ & 6.7 & 5.4 & 208 \\
Kerr & 0.03 & (4,1,3) & $1.17\times10^{-5}$ & 27.89$^*$ & 3.6 & 3.9 & 154 \\
5D & 0.01 & (2,1,1) & $2.81\times10^{-5}$ & 2.04 & 49.1 & 58.1 & 142 \\
5D & 0.01 & (3,1,2) & $1.63\times10^{-5}$ & 5.33 & 18.8 & 34.0 & 92 \\
5D & 0.01 & (4,1,3) & $1.24\times10^{-5}$ & 9.37 & 10.7 & 26.1 & 68 \\
5D & 0.02 & (2,1,1) & $8.11\times10^{-6}$ & 11.08$^*$ & 9.0 & 6.3 & 239 \\
5D & 0.02 & (3,1,2) & $4.21\times10^{-6}$ & 32.07$^*$ & 3.1 & 3.4 & 154 \\
5D & 0.02 & (4,1,3) & $3.03\times10^{-6}$ & 59.36$^*$ & 1.7 & 2.5 & 114 \\
5D & 0.03 & (2,1,1) & $3.37\times10^{-6}$ & 31.76$^*$ & 3.1 & 1.7 & 314 \\
5D & 0.03 & (3,1,2) & $1.63\times10^{-6}$ & 98.55$^*$ & 1.0 & 0.8 & 202 \\
5D & 0.03 & (4,1,3) & $1.13\times10^{-6}$ & 188.97$^\dagger$ & 0.5 & 0.6 & 149 \\
\end{tabular}
\end{ruledtabular}
\end{table*}

Since the orbit is periodic with period $T$ in the coordinate time $t$, its signal consists of spectral lines at frequencies $f_k=k/T$. We characterize the signal by the Fourier coefficients of one period,
\begin{equation}\label{eq:ck}
c_k=\frac{1}{T}\int_0^Th(t)\,e^{-2\pi i f_kt}\,dt\, ,
\end{equation}
which we compute with the discrete Fourier transform and which do not depend on the observation time. If the signal is repeated $N_{\rm eff}$ times, its Fourier transform at the lines is $\tilde h(f_k)=T_{\rm eff}\,c_k$, and each line has the width $\Delta f=1/T_{\rm eff}$. The signal-to-noise ratio is then given by  
\begin{equation}\label{eq:snr}
\rho^2=4\sum_k\frac{|\tilde h_+(f_k)|^2+|\tilde h_\times(f_k)|^2}{S_n(f_k)}\,\Delta f=\sum_k\left(\frac{h_{c,k}}{h_n(f_k)}\right)^2\, ,
\end{equation}
where $S_n(f)$ is the noise spectral density of the detector, $h_n(f)=\sqrt{fS_n(f)}$, and
\begin{equation}\label{eq:hc}
h_{c,k}=2\sqrt{f_k\,\Delta f}\left(\left|\tilde h_+(f_k)\right|^2+\left|\tilde h_\times(f_k)\right|^2\right)^{1/2}\, ,
\end{equation}
is the characteristic strain of the line $k$. With this definition, $(h_{c,k}/h_n)^2$ directly gives the contribution of each line to $\rho^2$. We note that Eq.~(\ref{eq:snr}) does not include the angular response of the detector, which changes $\rho$ by a factor of order unity.

For LISA~\cite{Amaro-Seoane2017LISA} and Taiji~\cite{10.1093/nsr/nwx116} we use the analytic sensitivity curve of Ref.~\cite{Robson2019} with the arm lengths $2.5\times10^9$\,m and $3\times10^9$\,m and the optical metrology noise $15$\,pm and $8$\,pm, respectively, and for TianQin we use the curve of Ref.~\cite{TianQin2016}.

In Table~\ref{tab:rad}, we provide the estimated lifetimes of the periodic orbits. For all orbits, the radiation reaction moves the orbit toward the separatrix. For the Kerr BH without magnetic field, the orbits are relatively far from the separatrix and survive about one week or longer. The magnetic field brings the orbits closer to the separatrix and strongly shortens their lifetime, i.e., for $\beta M=0.02$ the $(2,1,1)$ orbit survives about $20$\,h around the Kerr BH and about $6$\,h around the $D=5$ dimensional BH. We can deduce that orbits with larger $z$ lie closer to the separatrix and survive for a shorter time. For the $D=5$  dimensional BH with $\beta M=0.03$, the $(3,1,2)$ and $(4,1,3)$ orbits survive about one period or less, and for them our estimate is rough.

Consequently, for a stellar-mass compact object, the periodic orbits considered here describe short phases of the motion near the separatrix rather than long-lived sources. This is especially true for the $D=5$ dimensional BH, whose orbits lie closer to the separatrix.\\

\begin{figure*}[t]
\centering
\includegraphics[width=\textwidth]{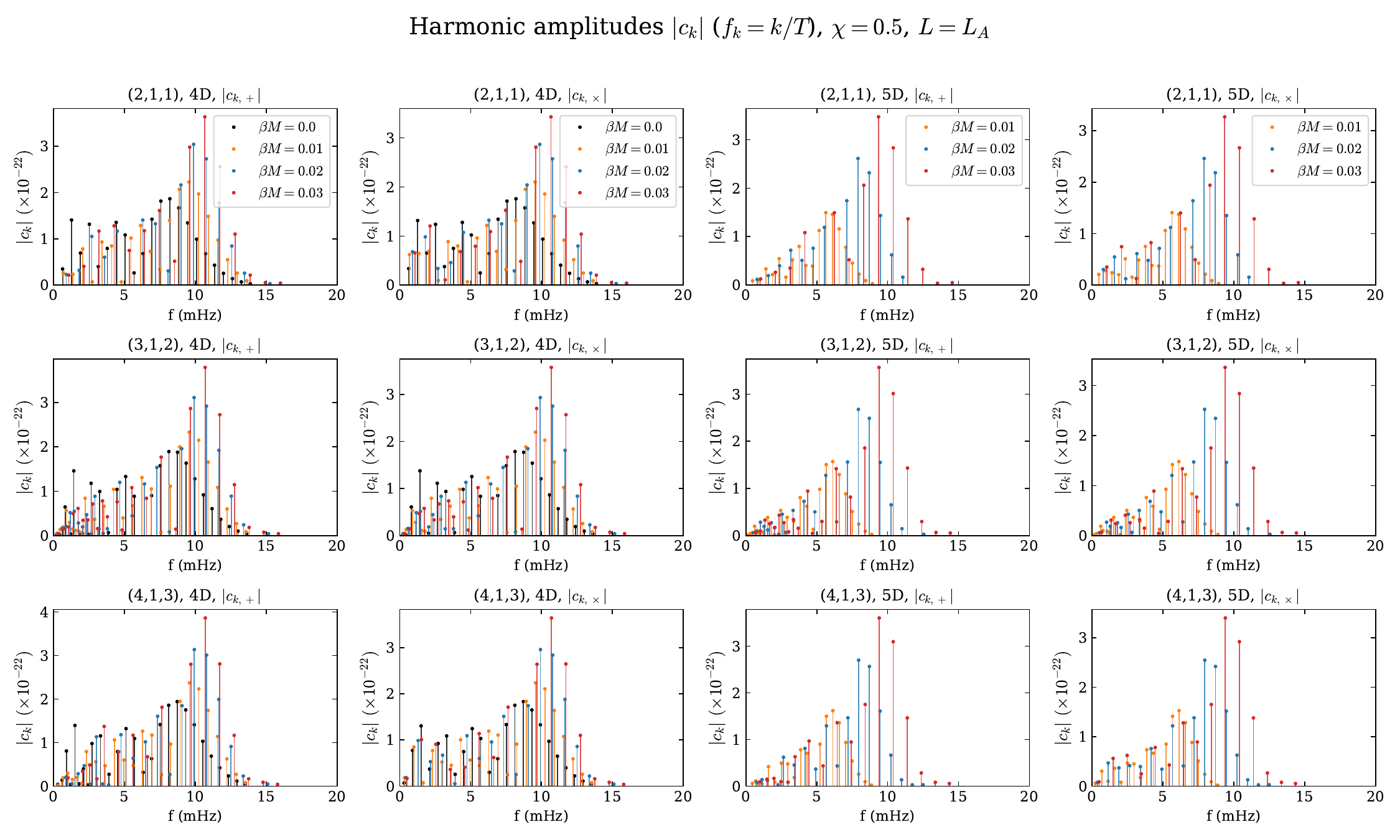}
\caption{Amplitudes $|c_k|$ of the spectral lines $f_k=k/T$ of the polarizations $h_+$ and $h_\times$ for the Kerr BH and the five-dimensional BH with $\chi=0.5$ and $L=L_A$. Lines weaker than $1\%$ of the strongest line are not shown.
}
\label{fig:spec}
\end{figure*}

\textbf{\textit{Spectra and detectability:}} In Fig~\ref{fig:spec} we show the amplitudes $|c_k|$ of the spectral lines. As shown in Fig~\ref{fig:spec}, the strongest lines are emitted during the whirls and lie at $5$--$11$\,mHz, within the sensitive band of the space-based detectors. For both Kerr and $D=5$ dimensional BHs, the spectrum moves to higher frequencies as $\beta$ increases. The spectrum of the $D=5$  dimensional BH lies at lower frequencies for the same $\beta$ and is weaker than that of the Kerr BH case.

\begin{figure*}[t]
\centering
\includegraphics[width=0.85\textwidth]{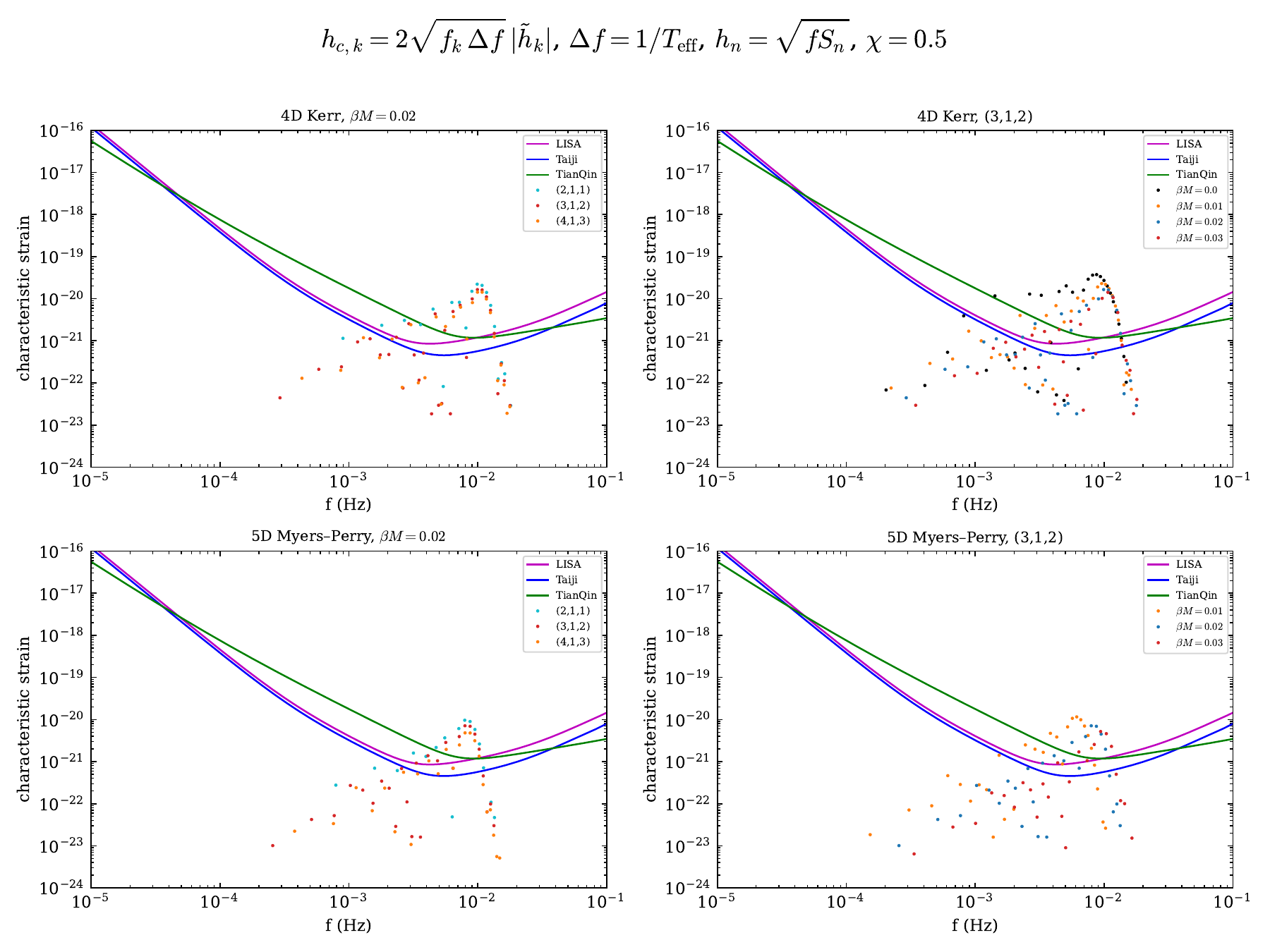}
\caption{Line strain $h_{c,k}$ of Eq.~(\ref{eq:hc}) for the effective observation time $T_{\rm eff}$ of each orbit, compared with $h_n=\sqrt{fS_n}$ of LISA, Taiji, and TianQin, for $\chi=0.5$. 
The squared ratio $(h_{c,k}/h_n)^2$ summed over the lines gives $\rho^2$. Note that each point corresponds to one spectral line at $f_k=k/T$.}
\label{fig:strain}
\end{figure*}

In Fig.~\ref{fig:strain}, we compare the line strain of Eq.~(\ref{eq:hc}) with the sensitivity curves, and we correspondingly provide the signal-to-noise ratios accumulated during the lifetime $T_{\rm eff}$ of each orbit in Table~\ref{tab:gw}. For the Kerr BH without magnetic field, $T_{\rm eff}$ is about one week and $\rho_{\rm LISA}\simeq80$. With the magnetic field the orbits live shorter, and $\rho_{\rm LISA}$ decreases to $15$--$54$ for the Kerr BH and to $7$--$31$ for the $D=5$ dimensional BH. \cblue{As discussed in Sec.~\ref{sec:gw}, the values for the five-dimensional BH are illustrative.} The values for Taiji are about twice as large. If the orbits did not evolve during one week, $\rho$ would be larger by a factor of up to about $14$, as shown in the last column of Table~\ref{tab:gw}. 

\section{Conclusions}\label{sec:conc}

Higher-dimensional BHs provide a useful theoretical laboratory for studying how strong-field gravity depends on the number of spacetime dimensions~\cite{EmparanReall2008,2021PDU....3100758S,Eshtursunov2026}. In higher dimensions, the gravitational potential falls off faster and the rotating BH has a different ergoregion structure \cite{2022IJMPD..3150120D}, which changes the motion of particles near the BH, and therefore the GWs emitted by them. Motivated by this, we studied periodic orbits of charged particles around the $D=5$ dimensional rotating BH immersed in a uniform magnetic field and the gravitational waveforms emitted by these orbits in an EMRI system. To separate the effects of the $D=5$ dimension from those of the magnetic field effects, we compared all results with the Kerr BH surrounded by a uniform magnetic field.

Since neutral particles have no stable bound orbits around the $D=5$ dimensional BH~\cite{2022IJMPD..3150120D,Dadhich22a}, the bound and periodic orbits appear only because of the magnetic field, and the ISCO moves to infinity as $\beta\to0$.
A charged particle is much more weakly bound to the $D=5$ dimensional BH, i.e., its ISCO energy is close to unity, $E_{\rm ISCO}\gtrsim0.95$, while for the Kerr BH it is noticeably smaller, $E_{\rm ISCO}\simeq0.8$--$0.94$. The region of bound orbits is much narrower, and the depth of the potential well is several times smaller for the $D=5$ dimensional BH (see Figs. \ref{fig:veff}, \ref{fig:isco} and \ref{fig:EL}). The ISCO of the Kerr BH depends only weakly on the magnetic field parameter, whereas for the $D=5$ dimensional BH its position is controlled by the magnetic field parameter, and its larger values push it to smaller radii (see Fig.~\ref{fig:isco}).

We found that the magnetic field lowers the minimal value $q_0$ of the rational number and can make it negative, i.e., $\omega_\phi<\omega_r$, which reflects the Larmor motion of the charge (see Figs. \ref{fig:qE} and \ref{fig:qL}). For the $D=5$ dimensional BH, the periodic orbits lie much closer to the separatrix than for the Kerr BH. Far from the BH the magnetic field force produces loops in the orbits, which are clearly seen for the Kerr BH and are small for the $D=5$ dimensional BH (see Figs. \ref{fig:orbL05} and \ref{fig:orbE05}). 

We showed that the GW signals of these orbits lie in the millihertz band. As $\beta$ increases, the orbital period decreases and the strongest spectral lines move to higher frequencies, and this effect is stronger for the five-dimensional BH. At the same magnetic field parameter, the $D=5$ dimensional signal has a longer period, lower frequencies, and a smaller amplitude than the signal for the Kerr BH case. This is because in the $D=5$ dimensional case the particle performs its whirls at a larger distance from the BH, where both the orbital frequency and the GW amplitude are lower (see Figs. \ref{fig:wf4D}, \ref{fig:wf5D}, ~\ref{fig:wfcomp} and Table~\ref{tab:gw}). For $\beta M=0.02$ the strongest line is at $7.9$\,mHz, as compared to $9.9$\,mHz for the Kerr BH, and the amplitude is about $18\%$ smaller. These differences decrease as the magnetic field parameter grows (see Fig. \ref{fig:spec} and Table \ref{tab:gw}).

We also estimated the effect of the radiation reaction. Our findings show that it moves the orbits toward the separatrix. For a compact object with $m=10M_\odot$, the periodic orbits with $\beta\neq0$ survive from less than an hour to a few days. For $\beta M\ge0.02$, orbits around the $D=5$ dimensional BH live several times shorter than the Kerr BH orbits. Over these lifetimes the signal-to-noise ratio for LISA is $15$--$54$ for the Kerr BH and $7$--$31$ for the $D=5$ dimensional BH, and about twice as large for Taiji (see Table \ref{tab:gw}). Therefore, the periodic orbits describe short phases of the motion near the separatrix. The signal-to-noise ratio must be computed over their lifetime rather than over a fixed observation time.

Finally, we computed the gravitational waveforms and provided an illustrative assessment of how orbital dynamics in the $D=5$ dimensional BH spacetime shaped the GW signal. However, we used a four-dimensional quadrupole formula, so the detectability estimates should be treated as illustrative. We estimated radiation reaction only to first order and omitted both orbital evolution through the separatrix and electromagnetic radiation from the charged body. We considered only the motion in the equatorial plane and did not study its stability with respect to perturbations out of the plane. In addition, we considered only the case $\beta L>0$, in which the Lorentz force points away from the BH. We consider these questions in our future work.

\begin{acknowledgments}

The authors wish to thank T. Zhu for useful discussions and comments.

\end{acknowledgments}

\bibliographystyle{apsrev4-1}
\bibliography{ref,new_refs}

\end{document}